\documentclass[11pt,notitlepage,aps,preprintnumbers,superscriptaddress,nofootinbib,aps,prd,floatfix]{revtex4-1}

\pdfoutput=1
\usepackage[utf8]{inputenc}
\usepackage[T1]{fontenc}
\usepackage{graphicx}
\usepackage{dcolumn}
\usepackage{bm}
\usepackage{slashed}
\usepackage{extarrows}
\usepackage{multirow,array}
\usepackage{amsmath,amssymb,physics}
\usepackage{enumitem}
\setlist{nolistsep}
\usepackage{xcolor}
\usepackage{url}
\usepackage{hyperref}
\usepackage{xspace}
\usepackage{orcidlink}
\definecolor{nicered}{rgb}{0.5,0.,0.}
\definecolor{nicegreen}{rgb}{0.,0.5,0.}
\definecolor{niceblue}{rgb}{0.,0.,0.5}
\hypersetup{colorlinks,citecolor=nicegreen,linkcolor=nicered,urlcolor=niceblue}
\usepackage[normalem]{ulem}

\usepackage{listings}
\usepackage[caption=false]{subfig}
\usepackage{ragged2e}
\usepackage{braket}
\usepackage{colortbl}
\usepackage{gensymb}

\usepackage[T1]{fontenc}

\newcommand{\GeV}{\textrm{GeV}}
\newcommand{\TeV}{\textrm{TeV}}

\usepackage[normalem]{ulem}

\newcommand{\whizard}{\texttt{Whizard}\xspace}

\begin{document}
	\setlength{\abovedisplayskip}{0pt}
	\setlength{\belowdisplayskip}{0pt}
	
	\title{Electroweak Parton Distribution Functions for Future Lepton Colliders}

\author{Tao Han\orcidlink{0000-0002-5543-0716}}
\affiliation{Pittsburgh Particle Physics Astrophysics and Cosmology Center,
Department of Physics and Astronomy,
University of Pittsburgh, Pittsburgh, PA 15260, USA\looseness=-1}
\author{Wolfgang Kilian\orcidlink{0000-0001-5521-5277}} 
\affiliation{Center for Particle Physics Siegen, University of Siegen, Walter-Flex-Str. 3, 57072 Siegen, Germany\looseness=-1}
\author{Yang Ma\orcidlink{0000-0002-9419-6598}}
\affiliation{Center for Cosmology, Particle Physics and Phenomenology (CP3), UCLouvain, B-1348 Louvain-la-Neuve, Belgium\looseness=-1}
\author{Krzysztof~M\k{e}ka{\l}a\orcidlink{0000-0003-4268-508X}}
\affiliation{Center for Particle Physics Siegen, University of Siegen, Walter-Flex-Str. 3, 57072 Siegen, Germany\looseness=-1}
\affiliation{Faculty of Physics, University of Warsaw, Pasteura 5, 02-093 Warsaw, Poland\looseness=-1}
\affiliation{Deutsches Elektronen-Synchrotron DESY, Notkestr. 85, 22607 Hamburg, Germany\looseness=-1}
\author{J\"urgen Reuter\orcidlink{0000-0003-1866-0157}}
\affiliation{Deutsches Elektronen-Synchrotron DESY, Notkestr. 85, 22607 Hamburg, Germany\looseness=-1}
\author{Keping Xie\orcidlink{0000-0003-4261-3393}}
\affiliation{State Key Laboratory of Dark Matter Physics \& Key Laboratory for Particle Astrophysics and Cosmology (MOE) \& Shanghai Key Laboratory for Particle Physics and Cosmology, Tsung-Dao Lee Institute \& School of Physics and Astronomy, Shanghai Jiao Tong University, Shanghai 201210, China\looseness=-1}
\affiliation{Department of Physics and Astronomy, Michigan State University, East Lansing, MI 48824, USA}

\date{\today}
	
\preprint{DESY-26-141, COMETA-2026-43, IRMP-CP3-26-34}   
\preprint{MLQC4FC-2026-0007, PITT-PACC-2619, SI-HEP 2026-24}
	
\begin{abstract}
At future high-energy colliders, such as a multi-TeV muon collider or the high-energy stages of an $e^+e^-$ linear collider, collinear radiation of electroweak gauge bosons becomes a dominant phenomenon. In this regime, many key processes are primarily initiated by photons and gauge bosons originating from lepton splittings. Electroweak Parton Distribution Functions (EW PDFs) provide an appropriate framework to describe these interactions by resumming large collinear logarithms arising from all possible Standard Model splittings, thereby enabling a partonic interpretation of the hard scattering.
The formalism is particularly suitable for studying new physics beyond the Standard Model in the electroweak sector at higher scales.
In this paper, we present the implementation of the EW PDF formalism within the Monte Carlo event generator \texttt{Whizard}. We demonstrate several applications of this framework and analyze its interplay with full fixed-order calculations as well as the purely kinematic Effective Weak Boson Approximation. We highlight scenarios where the EW PDF approach is especially useful for phenomenology and physics-potential studies for future lepton colliders. The EW PDF formalism is a vital step towards the ``electroweak jet regime'' where electroweak particles behave similarly to QCD jets. 
	\end{abstract}
	
	\maketitle
	\tableofcontents{}

\section{Introduction}
\label{sec:intro}

It is commonly accepted that the observed electromagnetic and weak interactions can effectively be described by a gauge theory of massless chiral fermions, scalar and vector bosons.  At a low temperature below the few-hundred $\GeV$ energy scale, they acquire masses via the Higgs mechanism of electroweak symmetry breaking.  If this theory, the Standard Model (SM), is correct at a more fundamental level, at high energies a phase of unbroken electroweak gauge symmetry with massless particles and jet-like multi-particle states is expected, which is analogous to and extends the familiar QCD phenomenology of collinear splitting and parton showers. 

However, the paradigm of electroweak symmetry restoration has so far been inaccessible to direct experimental verification. Electroweak jets should emerge for $E>v/\alpha$, where $v=246\;\GeV$ is the SM Higgs-field expectation value, and $\alpha$ denotes a typical gauge-coupling parameter.  Given the known electromagnetic and weak couplings, this amounts to $E\gtrsim$ 6--30\;\TeV.  Such energies are beyond the parton-level reach of the Large Hadron Collider (LHC).  Furthermore, electroweak precision data, which could receive corrections from high-mass new-physics effects indirectly, are generally sensitive only up to a few $\TeV$.  Hence, they do not yet probe the regime of the symmetric phase of the SM. 

Various collider concepts have been proposed which give direct access to partonic energy scales around $10\;\TeV$.  For instance, there are hadron-collider (FCC-hh~\cite{FCC:2018vvp}) as well as lepton-collider (the Muon Collider~\cite{Accettura:2023ked,InternationalMuonCollider:2025sys} or plasma-wakefield collider~\cite{Gessner:2025acq}) proposals.  At this scale, SM hard-scattering matrix elements will approximately factorize in the same way as hard processes factorize from long-range QCD in LHC processes.  The framework of parton-distribution functions (PDF) extends to all elementary degrees of freedom of the SM.  

The theory of electroweak PDFs (EW PDFs) was pioneered in the form of the Effective Weak Boson Approximation (EWA) for the $W$ and $Z$ bosons~\cite{Cahn:1983ip, Dawson:1984gx, Kane:1984bb, Lindfors:1985yp,Gunion:1986gm,Kleiss:1986xp,Kunszt:1987tk,Abbasabadi:1988ja,Kuss:1995yv,Kuss:1996ww}. This approximation was originally considered as a shortcut for studying high-energy new physics \cite{Duncan:1985vj,Johnson:1987tj,Barger:1990py} at supercolliders, such as the SSC project~\cite{Jackson1986SSC}. Recently, there has been renewed interest in the EWA and the conditions for applying it to high-energy processes~\cite{Accomando:2006mc,Borel:2012by,Costantini:2020stv,Ruiz:2021tdt,Bigaran:2025rvb,Frixione:2025guf,Frixione:2025wsv,Dahlen:2025udl}.

Beyond the electroweak vector bosons, a complete EW PDF framework incorporates all particles of the SM, from massless gluons and photons up to the top quark~\cite{Ciafaloni:2000df,Bauer:2017isx,Han:2020uid, Han:2021kes}. In the initial state of a hard-scattering process, these particles can repeatedly radiate and split, and thus mix with each other. The quantitative framework of combining and properly resumming leading-logarithmic contributions is given by the Dokshitzer--Gribov--Lipatov--Altarelli--Parisi (DGLAP) equations~\cite{Gribov:1972ri, Altarelli:1977zs, Dokshitzer:1977sg}.  In~\cite{Han:2020uid,Han:2021kes,Garosi:2023bvq,Dittmaier:2025htf}, the leading-order electroweak splitting functions and DGLAP equations at the leading logarithm have been set up and solved for the complete SM.  In the current paper, we adopt this formalism and make it available for numerical calculations and event generation in the universal Monte Carlo framework \whizard~\cite{Kilian:2007gr}. Its implementation in an MC generator allows the EW PDF formalism to be combined directly with other simulation stages available in \whizard.  For realistic event samples which cover this high-energy range, the description may be further supplemented by electroweak radiation in an electroweak parton-shower approach \cite{Christiansen:2015yqa,Chien:2018ohd,Kleiss:2020rcg,Masouminia:2021kne,Masouminia:2026udz} with proper account of mass thresholds in the radiation pattern. 

Any approximation should be discussed together with its limitations. For the EWA, and correspondingly for the complete EW PDF formalism, all relevant energy scales have to be well above the mass scale of the $W$, $Z$, Higgs, and top particles.  This condition excludes beam-momentum fractions below $x= \mathcal{O}(m_W/\sqrt{s})$ \cite{Dawson:1984gx,Dahlen:2025udl}. For the same reason, differential distributions in the exact forward region $p_T\lesssim m_W$ cannot be properly described.  Below the electroweak scale, one has to revert to a reduced PDF spectrum of light particles only. Additionally, the approximations to kinematics and radiation come with scale uncertainties.  In the numerical results which we present below, we will account for the limitations of the EW PDF approach, and the inclusion of scale uncertainties plays an important role.

For concreteness, we focus on $\mu^+\mu^-$ collisions at $10\;\TeV$ as the prime example, although our treatments and conclusions will be equally applicable to a high-energy $e^+e^-$ collider.  For leptonic beams, a first-principles perturbative treatment and solution of the DGLAP equations is feasible.
In Sec.~\ref{sec:implement}, we present the structure of the EW PDF implementation in \whizard. The quark and gluon partons and the QCD parton-induced dijet production at a muon collider are discussed in Sec.~\ref{sec:QCED}. The weak-boson PDF is covered in Sec.~\ref{sec:EW}, where we present a comparison between the resummed EW PDF and the leading-order EWA and discuss the scale choice for some typical SM processes. In Sec.~\ref{sec:concl}, we summarize the most important aspects of this work.

\section{The EW PDFs and the Whizard Implementation}
\label{sec:implement}

\subsection{The EW PDF Framework}

At scales $Q\gg m_{Z}$, the EW PDF formalism extends the partonic
picture to all relevant SM degrees of freedom. To leading
power in the weak scale over $Q$, their evolution is organized according
to the unbroken gauge group
$G_{\rm SM}=SU(3)_c\times SU(2)_L\times U(1)_Y$
\cite{Bauer:2017isx,Chen:2016wkt}. Quarks carry color. Left-handed quarks and leptons form weak doublets,
while right-handed charged fermions are weak singlets. The gauge fields are the gluon $g$, the weak-isospin fields
$W^a$ ($a=1,2,3$), and the hypercharge field $B$. These representations
fix the allowed splittings and their group factors; in particular, the
chiral weak couplings require separate fermion and gauge-boson helicity
PDFs~\cite{Bauer:2018arx}. A complete SM treatment also contains the Higgs
doublet, whose Goldstone components correspond at high energies to
longitudinal weak bosons~\cite{Chen:2016wkt}; the scalar-sector
approximation used in the present grids is specified below.

For a charged-lepton beam $\ell=e,\mu$, this partonic structure is
generated perturbatively following Refs.~\cite{Han:2020uid,Han:2021kes}.
The leading-order input at $Q=m_\ell$ is
\begin{equation}
  f_{\ell/\ell}(x,m_\ell^2)=\delta(1-x),
  \qquad f_{i/\ell}(x,m_\ell^2)=0\quad(i\ne\ell),
\end{equation}
where $x$ is the parton momentum fraction. For an unpolarized beam, the two
initial lepton helicities each carry half of the valence distribution.
The scale dependence is governed by the generalized
DGLAP equations~\cite{Gribov:1972ri,Altarelli:1977zs,Dokshitzer:1977sg}. 
Suppressing the beam label, the coupled evolution reads schematically

\begin{equation}
  \frac{\partial f_i(x,Q^2)}{\partial\ln Q^2}
  =\sum_I\frac{\alpha_I(Q^2)}{2\pi}\sum_j
  \int_x^1\frac{dz}{z}\,P^I_{ij}(z,Q^2)\,
  f_j\!\left(\frac{x}{z},Q^2\right).
  \label{eq:ewpdf-dglap}
\end{equation}
Here $P^I_{ij}$ is the kernel for $j\to i+\cdots$, including the
appropriate SM charges and helicity factors. Above the electroweak matching scale,
the interaction sum contains the $SU(3)_c$, $SU(2)_L$ and $U(1)_Y$ gauge contributions, with couplings
$\alpha_s$, $\alpha_2=g^2/(4\pi)$ and $\alpha_1=g'^2/(4\pi)$,
respectively, as well as the mixed neutral-gauge contribution. A complete
SM evolution also includes Yukawa kernels~\cite{Bauer:2017isx}.
The real terms populate daughter partons, while the virtual terms account
for the corresponding depletion of the parent distributions. Thus,
repeated splittings couple the fermion and gauge-boson PDFs and
redistribute the beam momentum among species. Solving these equations
with leading-order kernels and running couplings resums the leading
collinear logarithms. Weak-boson mass cutoffs enter the kernels, and
uncanceled soft contributions in electroweak non-singlet channels also
produce double logarithms~\cite{Bauer:2017isx,Bauer:2018arx}.

The evolution proceeds through three scale regions separated by two matching scales, $\mu_{\rm QCD}$ and $\mu_{\rm EW}$: 
\begin{itemize}
    \item Below $\mu_{\rm QCD}$ (i.e. the lightest hadrons), only QED splittings are included: photon radiation from the beam lepton generates photons, whose subsequent splittings generate sea leptons and quarks.
    \item Starting at $\mu_{\rm QCD}$, these quark distributions supply the input for coupled QED--QCD evolution, which also generates gluons. The currently supported matching scale choices are $\mu_{\rm QCD}=0.5\,\mathrm{GeV}$ and $\mu_{\rm QCD}=0.7\,\mathrm{GeV}$~\cite{BuarqueFranzosi:2021wrv}.  QED and QCD evolve together up to $\mu_{\rm EW}=m_Z$; individual heavy-flavor PDF threshold corrections are neglected in this approximation~\cite{Han:2021kes}.
    \item Above $\mu_{\rm EW}$, the evolution includes the unbroken $SU(3)_c\times SU(2)_L\times U(1)_Y$ gauge interactions, with the weak bosons, neutrinos and top quark becoming active. Fermion chiralities and transverse gauge-boson helicities are resolved separately. In particular, the transverse neutral sector is matched coherently between the physical and gauge bases. 
\end{itemize}
At the matching scale $Q=\mu_{\rm EW}=m_Z$, $f_Z=f_{\gamma Z}=0$, whereas the photon input gives~\cite{Han:2020uid}
\begin{equation}
  \left(f_B,f_{W^3},f_{BW^3}\right)
  =\left(c_W^2,s_W^2,2c_Ws_W\right)f_\gamma,
\end{equation}
for each transverse helicity, with $s_W=\sin\theta_W$ and $c_W=\cos\theta_W$ the sine and cosine of the weak mixing angle. The mixed distribution retains neutral-boson interference during evolution. Soft weak-boson radiation is regulated by $z_{\max}=1-m_Z/Q$, accounting for the uncanceled electroweak double logarithms. Chiral interactions consequently generate polarized PDFs even from unpolarized beams. The present \whizard calculation does not use the mixed distributions in the hard-scattering convolution, but the components are present in the PDF interface.

The longitudinal $W$- and $Z$-boson distributions are included at leading order using the EWA~\cite{Han:2020uid,Han:2021kes}. At this accuracy they have no leading logarithmic dependence on the factorization scale. The Higgs PDF is omitted from the grids used here, although its identifier is present in the interface.

Numerically, the procedure of Ref.~\cite{Han:2021kes} solves the evolution
in Mellin space using a fourth-order Runge--Kutta algorithm and returns
to $x$ space by numerical Mellin inversion along a Talbot contour. The
unresolved endpoint of the valence-lepton distribution is represented by
a local $\delta(1-x)$ term whose coefficient is fixed by the momentum sum
rule, $\sum_i\int_0^1 dx\,x f_i(x,Q^2)=1$. The resulting distributions
are tabulated on $(x,Q)$ grids for each parton and chirality, and
\whizard obtains the PDFs at the requested factorization scale by
interpolation, through the modified LHAPDF~\cite{Whalley:2005nh,Buckley:2014ana} interface. 
The details of the numbering scheme are described in App.~\ref{app:partons}.

\subsection{Implementation in \whizard}
\whizard\ is a multi-purpose Monte Carlo event generator package for all types of lepton and hadron colliders~\cite{Kilian:2007gr}. It provides leading-order matrix elements internally~\cite{Moretti:2001zz} and fully automates next-to-leading order QCD, EW and mixed corrections~\cite{Bredt:2022dmm,Campbell:2022qmc}, using external one-loop providers.

For the calculation and simulation of factorizable high-energy processes, \whizard\ generically implements iterative initial-state splitting in the collinear limit with convolution of beam spectra and PDFs.  For instance, leptonic collisions are described by the correlated beamstrahlung spectrum associated with a particular collider configuration~\cite{Ohl:1996fi}, convoluted with the universal PDFs which describe resummed QED initial-state radiation (ISR) \cite{Skrzypek:1990qs,Cacciari:1992pz,Frixione:2019lga,Bertone:2019hks}, and with the hard leptonic scattering matrix element.  In the analogous setup for hadron collisions, the program supports QCD-parton PDFs either natively or via the standard LHAPDF~\cite{Whalley:2005nh,Buckley:2014ana} interface.

For the purpose of event generation, the program supplements the partonic final state of the hard scattering by adding beam-remnant pseudoparticles which balance the conservation of energy-momentum, flavor, and color quantum numbers.  While in the lepton-ISR case, these pseudoparticles are equivalent to photons, hadronic collisions require a specific set of colored beam-remnant flavors.  The partonic final state combined with the beam-remnant configuration serves as input to standard QCD-shower and hadronization modules (e.g.~\cite{Kilian:2011ka,Kilian:2012pz}). 

In the current paper, we describe a new implementation of PDFs in \whizard\ where all SM particles can be treated as massless partons in the initial-state splitting kinematics.  The implementation builds upon the same generic framework as the PDF convolution chains described above.  The pre-generated grids described above are accessed by numerical interpolation, while the internal conventions and external user interface are analogous to LHAPDF.

In the electroweak factorization approach, beam remnants need additional consideration.  If all radiated particles are combined into pseudoparticle states analogous to QCD, various quantum-number combinations arise which cannot be handled yet by common parton-shower algorithms.  The results of this paper refer to final states at the parton level only, and the treatment of beam remnants and its connection to shower and hadronization is left for a future publication.

While collinear factorization is properly defined for massless particles, in practice partonic scattering matrix elements are evaluated for on-shell states.  This is not a problem for LHC processes where gluons, photons, and light quarks act as partons in high-energy collisions.  However, in the EW PDF approach, all SM particles are represented as massless partons with definite helicity while hard matrix elements are evaluated for massive particles.  Any factorization scheme thus incorporates a prescription for projecting massless partons onto their respective mass shell.  \whizard\ supports two distinct schemes where either energy or momentum is preserved in the overall splitting kinematics; in the following, the former is taken as default. 

The mapping of helicity states onto the polarization modes of massive particles is frame-dependent.  The default setting of \whizard\ refers to the laboratory frame, using the beam directions as reference.  Alternatively, we may define polarization states in the hard-process center-of-mass frame.  The frames are related by a collinear boost.  A boost along the quantization direction does not mix the polarization modes of collinear partons, massive or massless.  In this work, we do not analyze final-state polarization, for which the frame of reference is actually relevant.  We have checked both helicity projection schemes and explicitly verified that the choice of reference frame does not have an effect on the numerical results of this paper. 

We also note that there are significant contributions from Feynman diagrams which do not fit collinear-factorization kinematics and should be computed separately, e.g., fermion annihilation and subsequent decay.  We will discuss these issues and the range of validity of the EW PDF approach for various examples of our numeric results.

As a natural approach to massive particles in the EW PDF context, mass thresholds can consistently be accounted for in the PDF scale evolution, so if a particle mass is larger than the evolution scale, that particle is no longer considered a collinear parton.  The \whizard\ implementation could in principle distinguish all mass thresholds of the electroweak spectrum, namely the top-quark mass, the Higgs-boson mass, the $Z$-boson mass, and the $W$-boson mass.  However, the EW PDF grids from~\cite{Han:2020uid} which we use in the present study only support a single threshold, which is identified with the $Z$-boson mass. The $W$, $Z$, and Higgs bosons, as well as the top quark, are treated with their physical on-shell masses in the hard-scattering matrix elements. The $W$, $Z$, and top-quark PDF components become active above $m_Z$, while the Higgs PDF is omitted from the grids used here and its identifier is reserved in the interface. The treatment of the individual mass thresholds will be refined in future versions.

The \whizard\ user interface allows specifying a minimal value of Bjorken $x$ for either beam.  Since partons are treated as massive for the hard-process calculation, only values $x > m_\text{parton}/\sqrt{s}$ are physically meaningful.  Furthermore, users can additionally set the process factorization scale, either as a constant (e.g., related to the nominal collision energy) or as a dynamical expression (e.g., the event-dependent hard-process scale), as discussed later.  The factorization scale is an integral parameter of any scheme; it represents the phase-space boundaries which would apply to the full non-collinear process, and it controls the pattern of splitting and resummation in the evolution via the DGLAP equations within the factorized approximation.

The EW PDF implementation which we describe in this paper will be made available in an official release version of \whizard. Example \texttt{SINDARIN} input files for running EW PDFs in \whizard\ are provided in Appendix~\ref{app:sinEWPDF}.


\section{Processes in the PDF Framework below the Electroweak Scale}
\label{sec:QCED}

If some characteristic energy scales of a hard process lie below the electroweak mass threshold $\mu_{\rm EW}$ (chosen equal to $m_Z$), the heavy electroweak particles $W,\,Z,\,H,\,t$ are not yet excited in the parton spectrum.  The remaining partons, which we treat as massless in the factorization approximation, are leptons, light quarks, the photon and the gluon. While colored partons obviously dominate the content of a hadron beam, in this section we will focus on lepton beams where QCD radiation is formally a subleading effect, but can lead to large cross sections as shown below. 

In the collinear factorization limit, the leading partonic content of a beam lepton is determined by photon emission.  Hard processes can be initiated either by the original lepton after having emitted an arbitrary number of photons (ISR), or by a photon radiated from the beam lepton.  In the latter case, the leading logarithmic behavior is captured by the traditional equivalent photon approximation (EPA)~\cite{Fermi:1924tc,vonWeizsacker:1934nji,Williams:1934ad}. In the EW PDF formalism~\cite{Han:2020uid, Han:2021kes}, the evolution of the PDF is governed by the DGLAP equations, which incorporate the full spectrum of light particles and thus unify the ISR and EPA approaches.  Multiple splittings introduce all light particles as partons of the beam lepton.  In all configurations, the large logarithms associated with repeated collinear splittings are resummed into the PDF values.
Following Ref.~\cite{Drees:1994eu}, we introduce a matching scale $\mu_{\rm QCD}$ which separates the QED- and QCD-dominated regimes. Below $\mu_{\rm QCD}$, the DGLAP evolution includes only QED interactions.
Above $\mu_{\rm QCD}$, QCD interactions are accounted for, and the gluon becomes an active parton. This matching scale parametrizes the non-perturbative threshold effects and, in principle, should be determined from experimental data or lattice simulations. The nominal matching scale is $\mu_{\rm QCD}=0.5$ GeV, with alternative PDF sets with $\mu_{\rm QCD}=0.7$ GeV provided to quantify the non-perturbative matching variation~\cite{BuarqueFranzosi:2021wrv}. For the numerical comparisons in this section, we use the $0.7$ GeV set unless stated otherwise. 

For concrete applications in this paper, we consider a multi-TeV muon collider~\cite{InternationalMuonCollider:2025sys} with an integrated luminosity assumed to scale as
\begin{equation}
    \mathcal{L}_\text{int} = \left( \frac{\sqrt{s}}{10\,\text{TeV}} \right)^2 \times 10\,\text{ab}^{-1},
\end{equation}
and take $\sqrt{s}=10$ TeV as a benchmark energy.  At a muon collider, QCD-induced dijet production provides an important contribution to the SM background and can be comparable to, or even larger than the photon-induced process $\gamma\gamma\to q\bar{q}$~\cite{Han:2021kes}. 
By including the photon, quark, and gluon PDFs of the muon, we consider the following partonic processes contributing to dijet production (we do not consider the contribution from the valence muons, $\mu^+\mu^-\to jj$, which is only important at the highest dijet invariant mass bins):
\begin{equation}
\begin{array}{rclcrccl}
\gamma\gamma &\to& qq & {}\quad{}
& \gamma g &\to qq &
{}\quad{}
& \gamma q \to gq
\\
qq &\to& qq\;(gg) & {}\quad{}
& gq &\to gq &
{}\quad{}
& gg \to gg\;(qq)
\end{array}
\end{equation}
where $q$ denotes both light quarks and antiquarks.
We choose the factorization scale as $Q=\sqrt{\hat{s}}/2$ and impose the following kinematic cuts: a minimum invariant mass of 5~GeV for the dijet system and a minimum transverse momentum of 1~GeV for each jet. For simplicity, detector effects are neglected, and the jet kinematics are taken directly from the final-state quarks produced in the generator-level hard process.

\begin{figure}[htb]
    \subfloat[all jets]{
		\centering
		\includegraphics[width=0.49\textwidth]{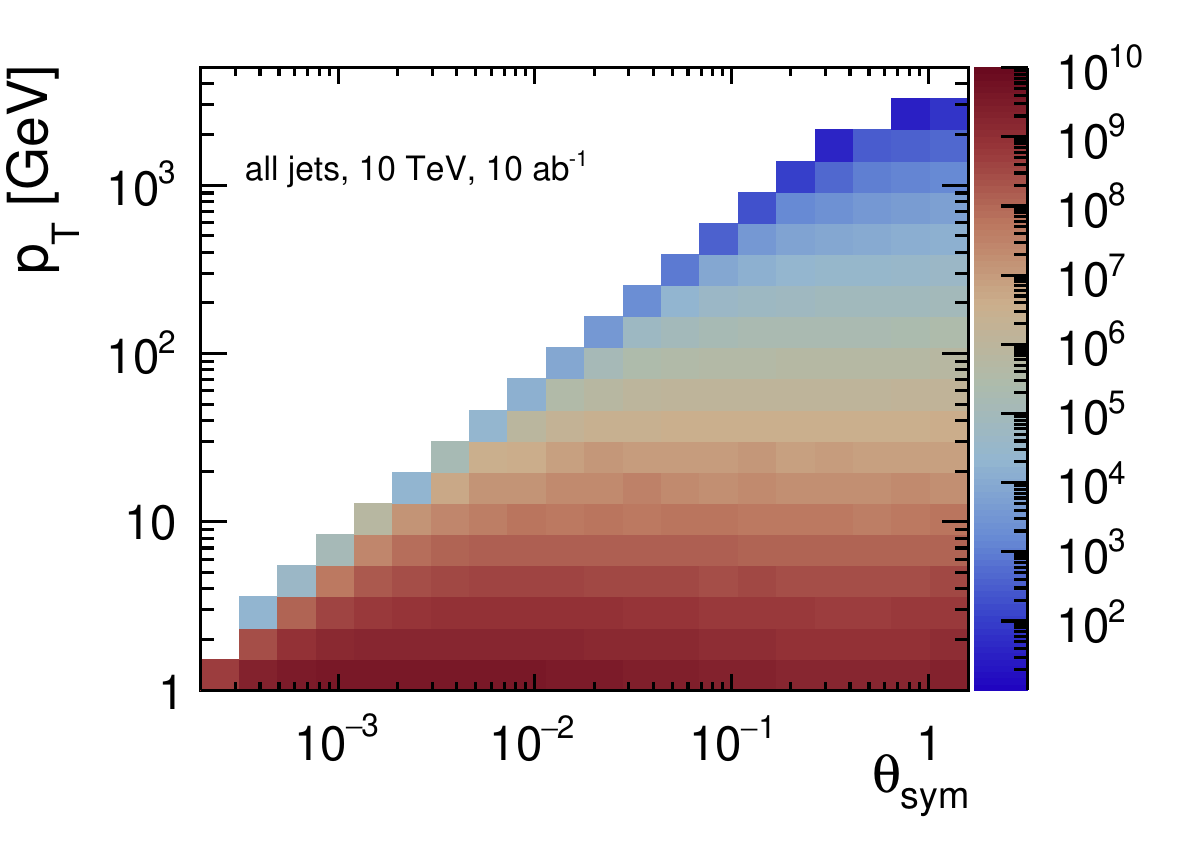}}
	\hfill
		\centering
        \subfloat[ratio of QCD-ind. and $\gamma\gamma$-ind. events]{
		\includegraphics[width=0.49\textwidth]{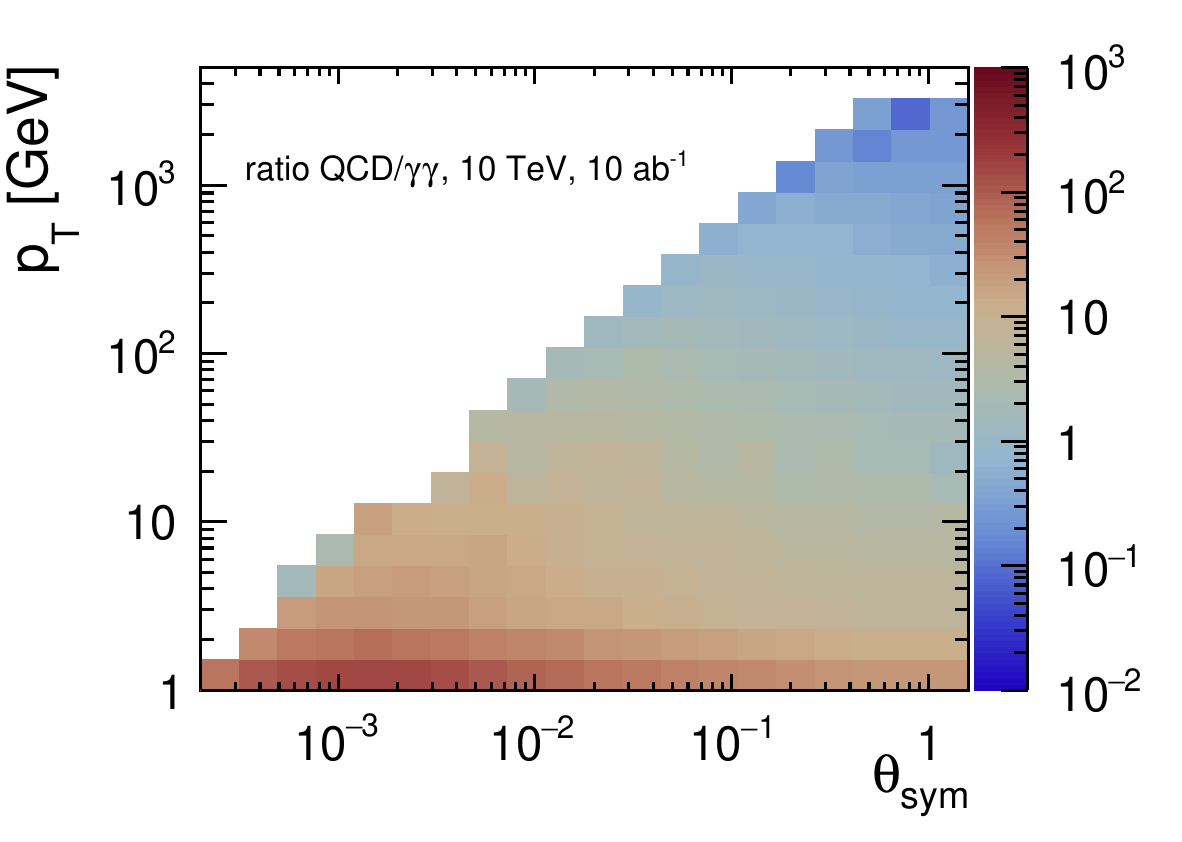}}
	\caption{Two-dimensional distributions of the transverse momenta and symmetrized emission angles of jets produced at a 10-TeV muon collider, assuming an integrated luminosity of 10\,ab$^{-1}$. The left plot shows jets from all the channels, while the right one shows the ratio of the colored-parton-initiated processes over the $\gamma\gamma$ contribution.}
    \label{fig:EWPDF_colored_2D}
\end{figure}

For illustration, Fig.~\ref{fig:EWPDF_colored_2D} shows the two-dimensional distribution in $(p_T^j,\theta_{\rm sym})$ at a 10-TeV muon collider. The symmetrized angle, $\theta_{\rm sym}$, is defined as the smaller angle between the jet and the beam axis, and is introduced to take advantage of the log scale on the x-axis. The left panel shows the sum of all the production channels, while the right panel shows the ratio of the processes initiated by colored partons over the photon-induced process $\gamma\gamma\to q\bar{q}$. The yields are obtained assuming $\mathcal{L}_{\rm int}=10$\,ab$^{-1}$. The QCD-induced contribution is found to be dominant in the soft region with $p_T^j<10$\,GeV.

\begin{figure}[htb]
	\centering
	\includegraphics[width=0.6\textwidth]{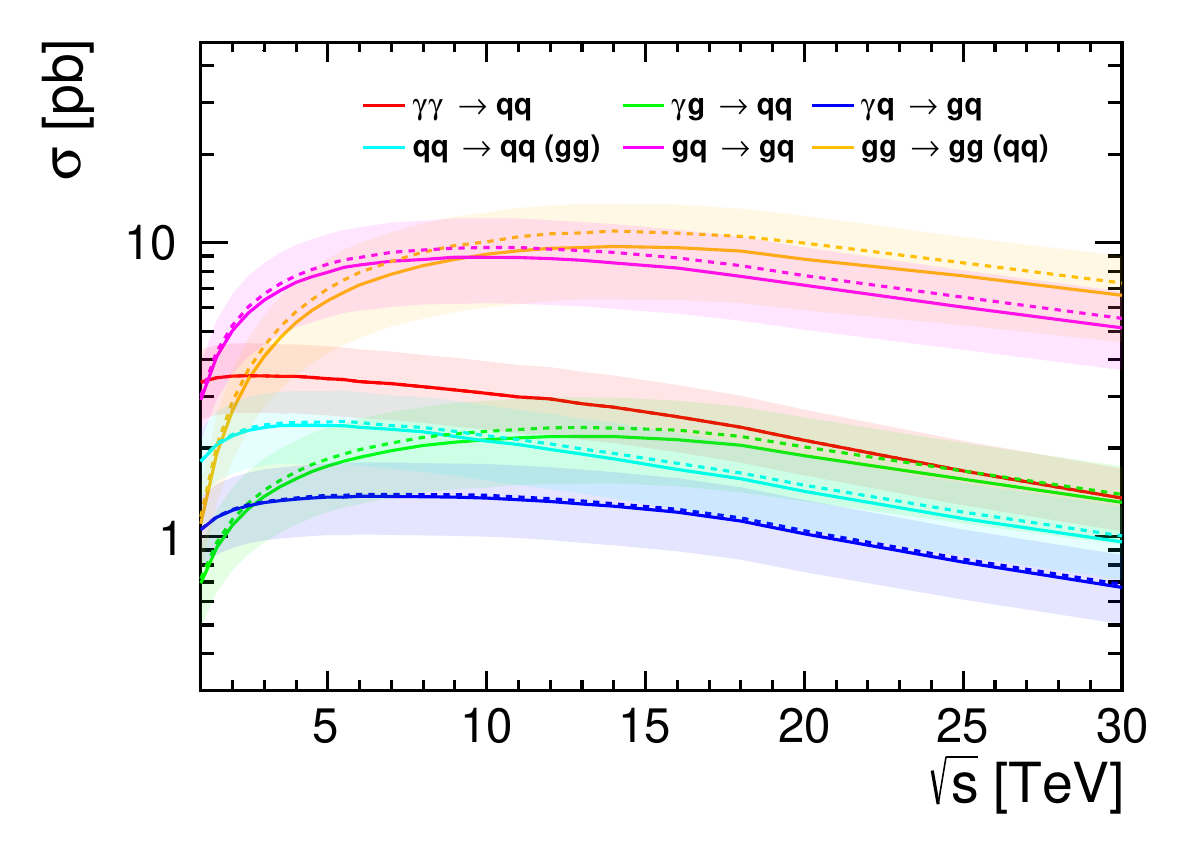}
	\caption{Cross section for various dijet production channels evaluated within the EW PDF framework for collision energies between 1 and 30 TeV using $Q = \sqrt{\hat{s}}/2$ and $\mu_{\rm QCD}=0.7$\,GeV. The colored bands show the factorization-scale variation by a factor of two in both directions. The dotted lines indicate the results for the central value of the scale assuming $\mu_{\rm QCD}=0.5$\,GeV. For more details cf. the main text.}
	\label{fig:EWPDF_colored_scale}
\end{figure}

In Fig.~\ref{fig:EWPDF_colored_scale}, we present the cross sections for the dijet production across a range of center-of-mass energies, again using a factorization scale $\sqrt{\hat{s}}/2$. Simple acceptance cuts are adopted following Ref.~\cite{Han:2021kes}:
\begin{equation}
p_T^j > \left(4 + {\sqrt s \over 3\ {\rm TeV}}\right) \,\GeV, \quad m_{jj}> 20 \,\GeV, \quad |\eta_j|<2.44.
\end{equation}
As previously emphasized in Refs.~\cite{Han:2021kes,Frixione:2023gmf}, the cross sections are sizable, of the order of picobarns. 

The main contributions stem from gluon--gluon and gluon--quark interactions, while other channels remain of comparable size, differing by at most one order of magnitude at the highest energies. We emphasize that the high-energy behavior comes from the energy-dependent $p_T^j$ cut. A less stringent one, for example, independent of the collision energy, would keep the contributions growing with energy. 
Our results are fully consistent with \cite{Han:2021kes}. Additionally, we verified that, under the same cuts, our results differ from those reported in Ref.~\cite{Frixione:2023gmf} by no more than 5--10\%.

Fig.~\ref{fig:EWPDF_colored_scale} uses the scale of $\sqrt{\hat{s}}/2$ and shows its variation by a factor of two up and down. The scale uncertainties are well under control and do not invalidate the conclusions regarding the size of the dijet contributions. The same applies to the variation of the QCD scale; as shown in the figure, the scale of 0.5\,GeV does not change the results significantly.

\section{Processes in the PDF Framework above the Electroweak Scale}
\label{sec:EW}

We now turn to the high-energy regime where, ideally, particle masses can be neglected altogether.  All SM particles belong to the spectrum of massless partons which mix with each other in the DGLAP evolution.  
Splittings involving heavy particles tend to favor low values of $x$ for the parton entering the hard process.
Multiple splittings easily result in $x\lesssim m_Z/\sqrt{s}$, such that the conditions for collinear factorization are no longer met. 
Taking a 10-TeV muon collider as our prime example, we mostly consider a single massive boson as an effective parton of each muon beam. 
This simplified picture of collinear factorization has been available for a long time as the EWA~\cite{Dawson:1984gx,Kane:1984bb}.  However, by incorporating the SM particles in a unified EW PDF framework, we extend the EWA and fully account for extra light radiation, and eventually also describe higher-order heavy radiation at asymptotic energies.  In effect, EW PDFs resum higher-order logarithmic corrections to the EWA and EPA results.

On the other hand, automated programs such as \whizard\ can calculate processes at fixed order, without higher-order radiation, using complete matrix elements and exact kinematics.  It is therefore meaningful to compare predictions obtained within either the EW PDF or EPA/EWA factorized approaches with fixed-order matrix-element results.   In the following, we will perform such a comparison for several example processes and comment on the differences and deviations.

As in the previous section, we focus on a 10-TeV muon collider and assume collecting 10 ab$^{-1}$ of data.  We present our new EW PDF results together with the corresponding fixed-order calculation and the EWA implementation in \whizard, first introduced in~\cite{Dahlen:2025udl}.  Regarding the latter, we choose the ``\texttt{log}'' version of the EWA formula which coincides with the leading-order piece of the EW PDF result.  The numerical results in Ref.~\cite{Dahlen:2025udl} were obtained using the ``\texttt{default}'' version, as explained in that paper.

For the different processes in the following subsections, we will show results for the lower and the upper half of the invariant mass range separately to disentangle effects from the threshold region and the matching to the prompt production for the full processes. As a general remark, it should be stressed that, within the EW PDF formalism, fully inclusive predictions generally require summing over all possible initial-state partons that can contribute to a given experimental signature. In the present study, the selection is restricted to specific sub-processes to enable a direct and transparent comparison between the EW PDF and EWA frameworks. These channels are also expected to yield the major contributions.

\subsection{Higgs pair production}
\begin{figure}
	\centering
		\subfloat[$m_{HH} < 5$\,TeV]{\includegraphics[width=0.49\textwidth]{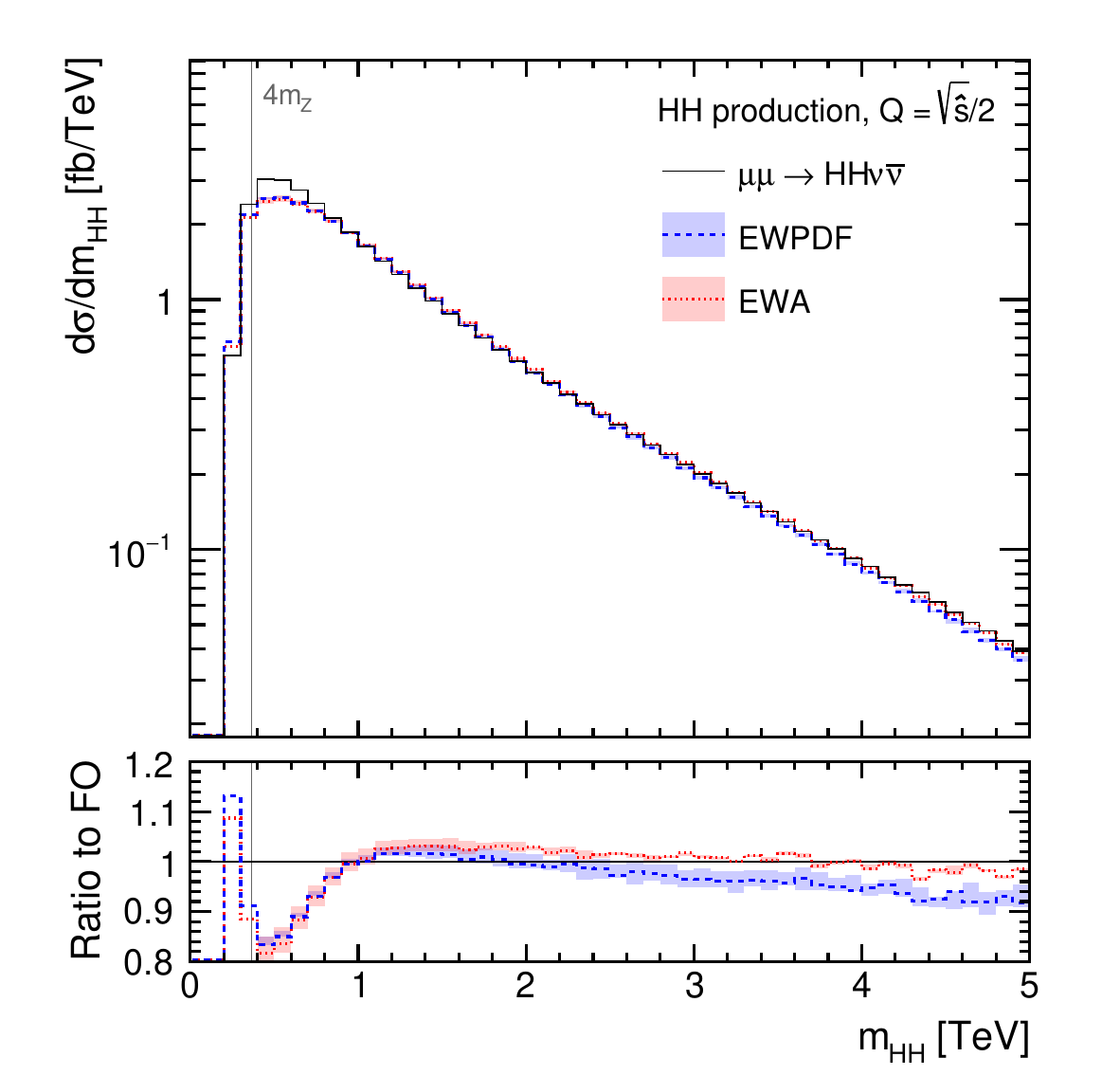}}
	\hfill
		\subfloat[$m_{HH} > 5$\,TeV]{\includegraphics[width=0.49\textwidth]{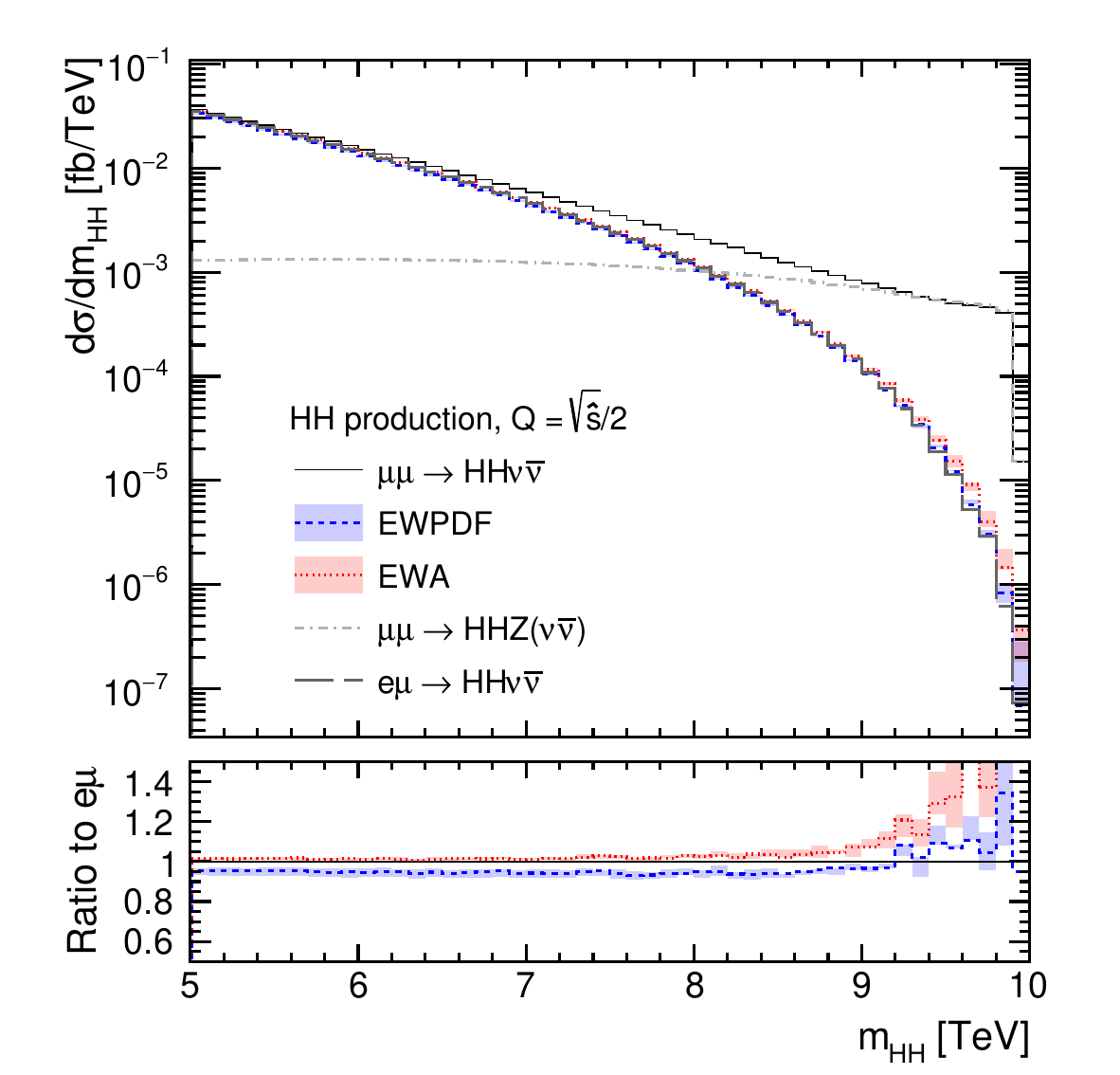}}
	\caption{Invariant mass distributions of the di-Higgs system in the process $\mu^- \mu^+ \to HH \nu_\mu\bar{\nu}_\mu$ at $\sqrt{s} = 10$\,TeV for the full matrix element calculation (black), the EW PDF (blue) and the EWA (red). Two additional lines are shown in the right-hand plot: dark gray for the full process $e^- \mu^+ \to HH\nu_e \bar{\nu}_\mu$ and dash-dotted light gray for the process $\mu^- \mu^+ \to ZHH$ with a subsequent decay $Z\to \nu_\mu \bar{\nu}_\mu$.
    The bands represent a variation of the scale by a factor of two and are shown only for bins above $4m_Z$. The left bottom panel shows the ratio to the fixed-order calculation, while the right panel shows the ratio to the fixed-order calculation with $e\mu$ beams to reduce non-VBF topologies.}
	\label{fig:EWPDF_HH}
\end{figure}

\begin{figure}
	\centering
		\subfloat[leading $p_T$]{
		\includegraphics[width=0.49\textwidth]{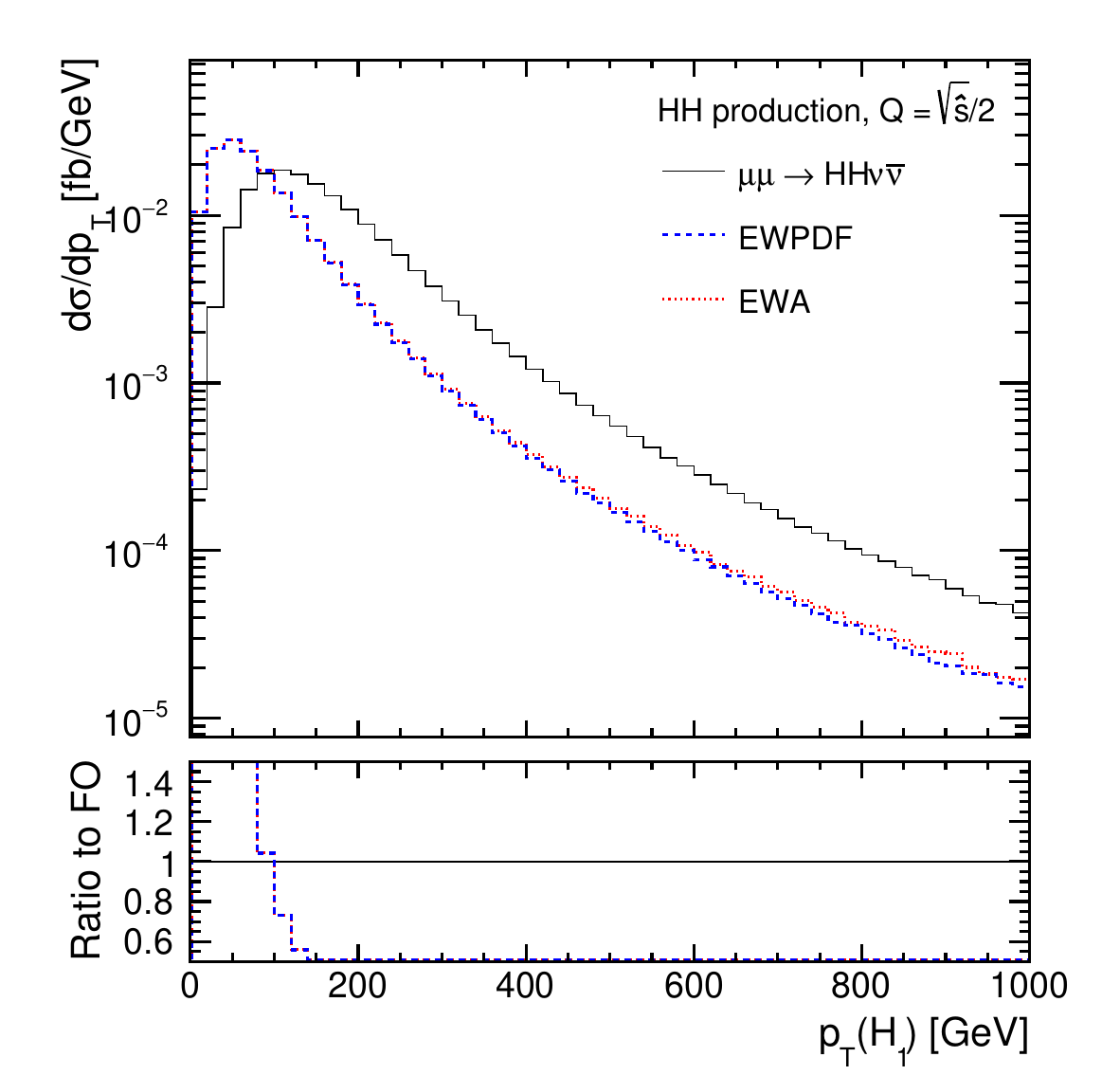}}
	\hfill
		\centering
		\subfloat[subleading $p_T$]{\includegraphics[width=0.49\textwidth]{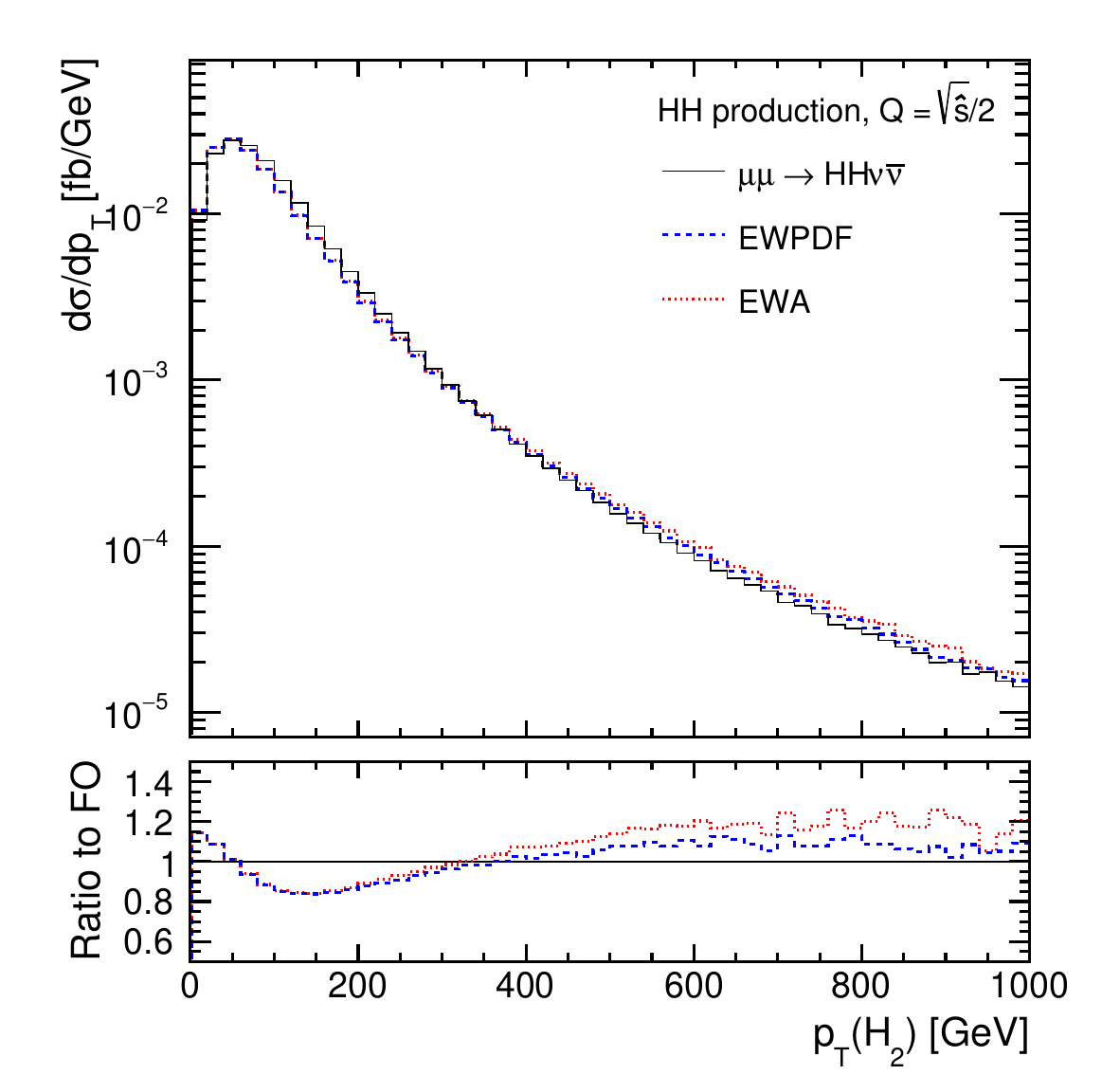}}
	\caption{Transverse momentum distributions of the leading (left panel) and subleading (right panel) Higgs bosons in the process $\mu^- \mu^+ \to HH\nu_\mu \bar{\nu}_\mu$ at $\sqrt{s} = 10$\,TeV for the full matrix element calculation (black), the EW PDF (blue) and the EWA (red). The bottom panels show the ratio to the fixed-order calculation.}
	\label{fig:EWPDF_HH_pT}
\end{figure}

\begin{figure}
	\centering
		\subfloat[invariant mass]{\includegraphics[width=0.49\textwidth]{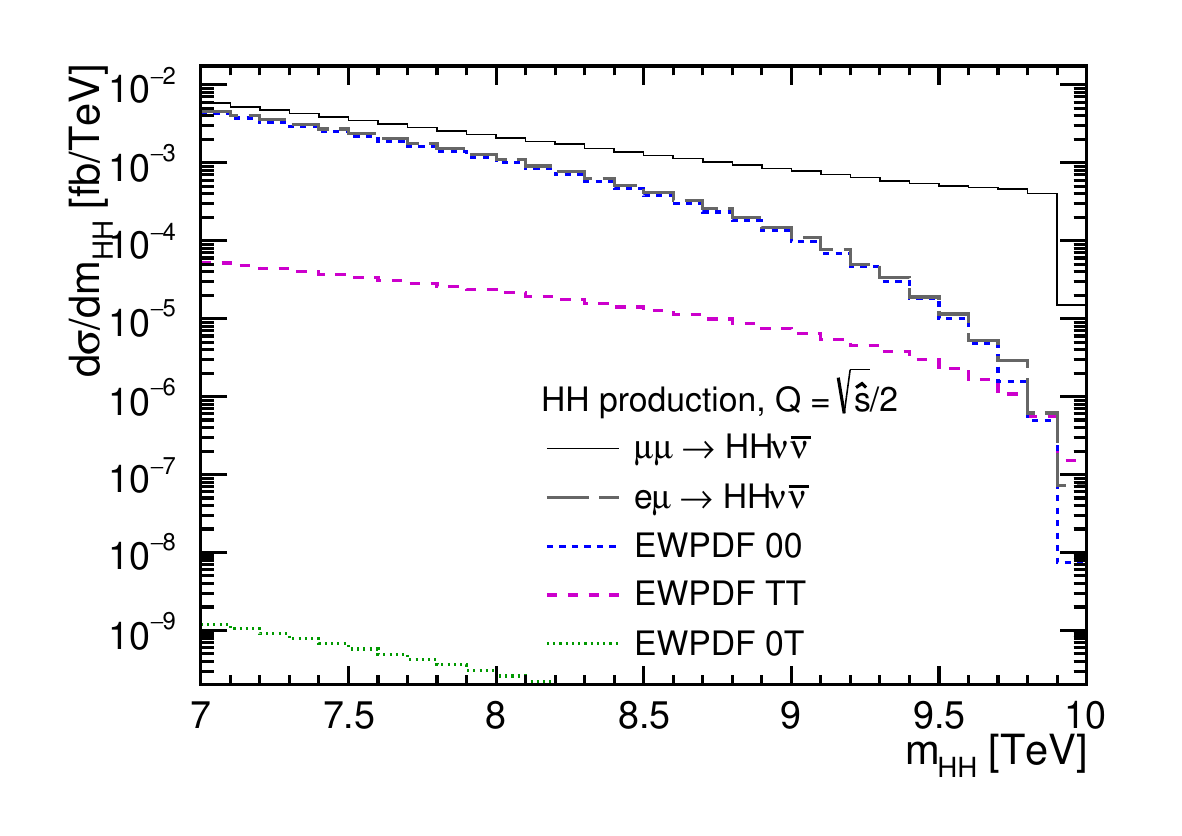}}
	\hfill
		\subfloat[subleading $p_T$]{\includegraphics[width=0.49\textwidth]{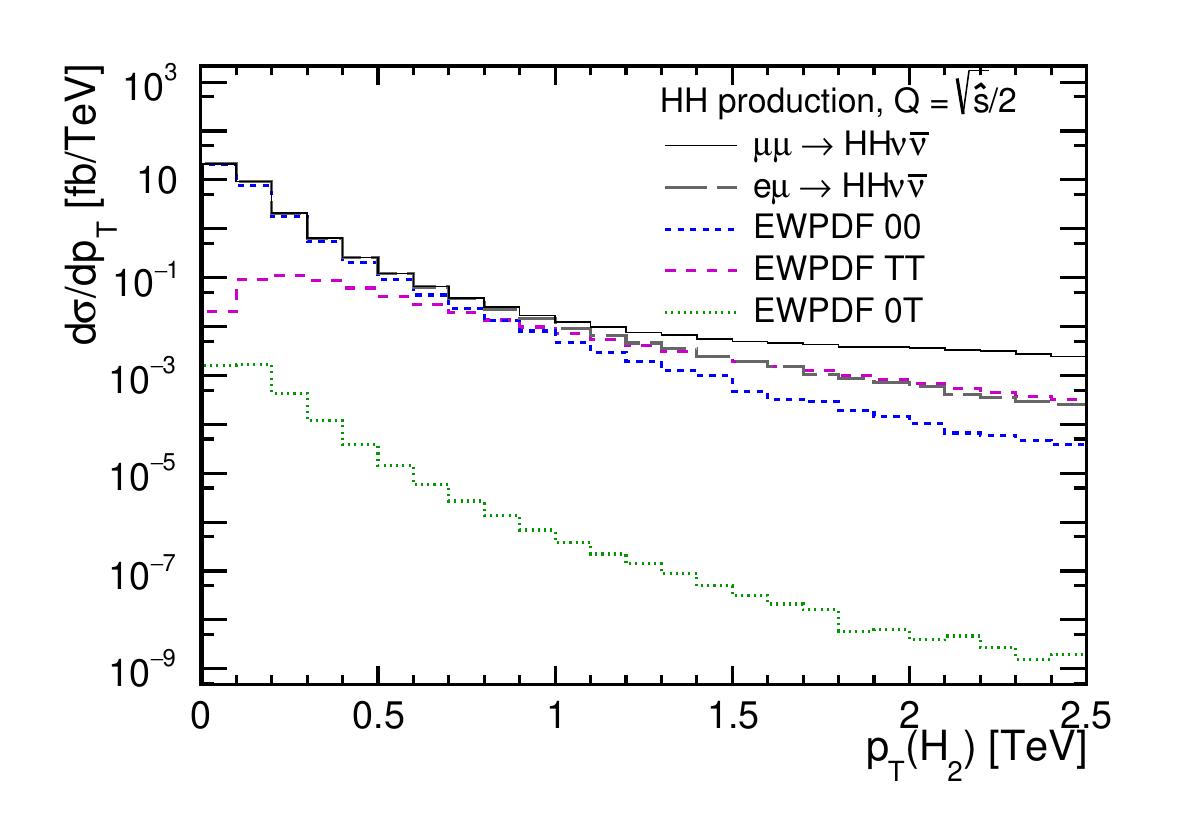}}
	\caption{Left: Invariant mass distributions of the $HH$ system in the process $\mu^- \mu^+ \to HH \nu_\mu\bar{\nu}_\mu$ at $\sqrt{s} = 10$\,TeV. Right: Transverse momentum distributions of the subleading partons in the same process. The full matrix element calculation is shown in black, the blue curve indicates the longitudinal contribution in the EW PDF framework, the magenta curve the transverse contribution, and the green curve the longitudinal-transverse contribution. An additional line is shown in gray to illustrate the full process $e^- \mu^+ \to HH\nu_e \bar{\nu}_\mu$.}
	\label{fig:EWPDF_HH_pol}
\end{figure}

First, we consider the process $\mu^- \mu^+ \to HH \nu_\mu \bar{\nu}_\mu$.  In the EW PDF framework, the hard process is $W^+W^-\to HH$, where all particles are considered to be massive.  To leading order, the $W$ bosons originate from $\mu\to W\nu$ splitting, where masses are neglected.  As explained above, the splitting results in pure-helicity states, which we match to on-shell polarization states of the $W$ bosons in hard processes. 

Fig.~\ref{fig:EWPDF_HH} shows the invariant mass distribution of the di-Higgs system, comparing the EWA and EW PDF predictions with the full matrix-element calculation\footnote{For notational simplicity, beam-remnant neutrinos are denoted without a flavor index in the plots. In the following, we consider only processes in which their flavor matches that of the corresponding beam.}.  For this particular process, varying the factorization scale has a negligible effect on the results, as the main contribution to the process is given by the longitudinal $W$ bosons, which are not logarithmically enhanced. The collinear approximation provides a very good description of the full process over most of the invariant-mass range.  At the lower end near the  $HH$ threshold (left plot), the on-shell projection becomes unreliable because $x\gg m_W/\sqrt{s}$ is no longer satisfied.  In this region, the factorized approximation undershoots the full matrix-element calculation. 

At the upper end of the spectrum (right plot), Feynman graphs with annihilation and thus non-VBF-like topology ($\mu^+\mu^-\to ZHH\to \nu_\mu\bar\nu _\mu HH$) come into play and eventually dominate the distribution.  However, their contribution to the total cross section is marginal.
For illustration, we can remove the annihilation contribution by replacing one of the beam muons by an electron, i.e., an $e^-\mu^+$ collider with the same parameters as the $\mu^+\mu^-$ collider.  This is displayed in Fig.~\ref{fig:EWPDF_HH}, right.
In fact, the corresponding curve agrees with the EWA and EW PDF predictions up to the phase-space limits.  

A close look at the right panel of Fig.~\ref{fig:EWPDF_HH} reveals that the EWA curve matches the full leading-order result perfectly as long as $m_{HH}$ does not approach either end of phase space.  This is expected since in that range, the kinematical approximations of the EWA are justified.  By contrast, the EW PDF curve exhibits a reduction of about $5\%$.  We argue that this effect is due to the higher-order resummation of radiation being accounted for in the EW PDF approach. This resummation results in the usual scale dependence of the effective coupling constant. It also reduces the differential cross section at the upper end of the phase space.  Both effects are meaningful corrections to the fixed-order result.  For the process at hand, this correction is small since di-Higgs production is dominated by longitudinal $W$ bosons. This behavior highlights a fundamental conceptual difference manifested at high-$x$: unlike the EWA framework, which --- as a kinematical approximation --- does not constrain the flux of effective bosons, the EW PDF framework preserves the momentum sum rules.

In Fig.~\ref{fig:EWPDF_HH_pT}, we present the transverse-momentum distribution of leading and subleading Higgs bosons.  As indicated in Ref.~\cite{Dahlen:2025udl}, the subleading parton tends to be better described within the realm of the collinear approximation due to the recoil effect for the leading parton in the full $2\to 4$ process. As follows from the figure, care must be taken when applying kinematical cuts; in general, as the kinematics of the $2\to 4$ and $2\to 2$ processes are different, the fully inclusive predictions tend to be more accurate.

Another feature of the EW PDF formalism is its ability to separate the different helicity components of the interaction, which are otherwise summed over in the full ME description. Fig.~\ref{fig:EWPDF_HH_pol} (left) shows the invariant mass distribution of the di-Higgs system at large invariant masses, $m_{HH} > 7$\,TeV. As expected, the process is dominated by longitudinally polarized $W$ bosons, except in the very last bins, where the transverse components become comparable in size. In Fig.~\ref{fig:EWPDF_HH_pol} (right), we show the subleading-Higgs $p_T$ distribution without any cut. In this case, the transverse components start to dominate already at about 1\,TeV, suggesting the possibility of a physical, observable-based separation of the underlying helicity components. This observation is particularly interesting, as it may provide new insight into the structure of electroweak interactions and offer a way to probe fundamental aspects of the Standard Model, such as the Goldstone Boson Equivalence Theorem~\cite{Lee:1977eg,Chanowitz:1985hj,Cuomo:2019siu}. We will investigate this phenomenon in a dedicated future study.

\subsection{Top quark pair production}
\begin{figure}
	\centering
		\subfloat[$m_{t\bar{t}} < 5$\,TeV]{\includegraphics[width=0.49\textwidth]{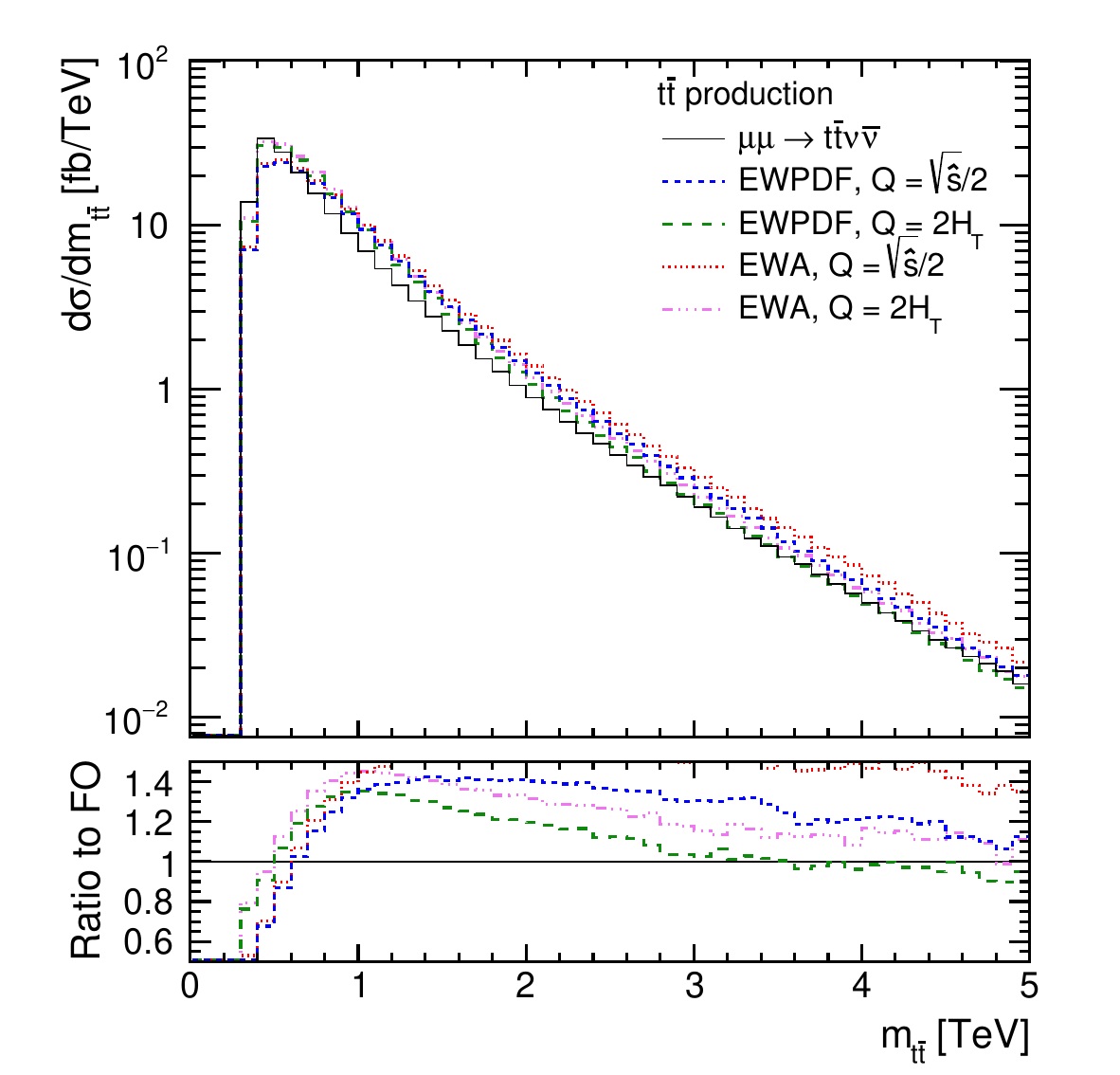}}
	\hfill
		\subfloat[$m_{t\bar{t}} > 5$\,TeV]{\includegraphics[width=0.49\textwidth]{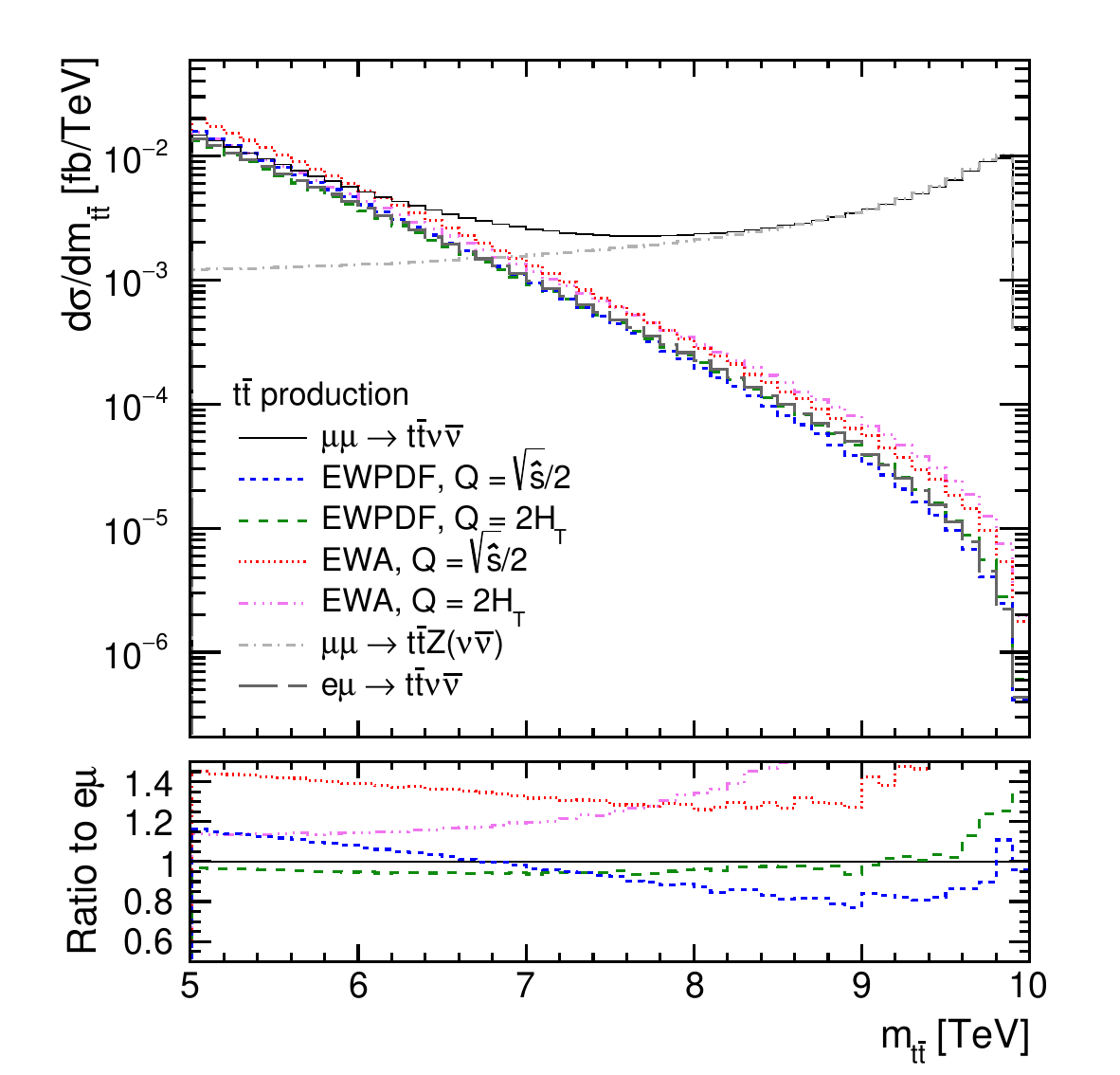}}
	\caption{Invariant mass distributions of the $t \bar{t}$-system produced in the $W^+W^-$ interaction at a $\mu^- \mu^+$ collider for the full matrix element calculation (black), the EW PDF and the EWA. The blue curve indicates the EW PDF calculation with a scale of $\sqrt{\hat{s}}/2$ and the green curve with a scale of $2H_{T}$; the red curve stands for the EWA calculation with a scale choice of $\sqrt{\hat{s}}/2$, while the pink one --- a scale of $2H_{T}$. Two additional lines are shown for high invariant masses: dark gray for the full process $e^- \mu^+ \to t\bar{t}\nu_e \bar{\nu}_\mu$ and dash-dotted light gray for the process $\mu^- \mu^+ \to Zt\bar{t}$ with a subsequent decay $Z\to \nu_\mu \bar{\nu}_\mu$. The left bottom panel shows the ratio to the fixed-order calculation, while the right panel shows the ratio to the fixed-order calculation with $e\mu$ beams to reduce non-VBF topologies.}
	\label{fig:EWPDF_ttbar}
\end{figure}

\begin{figure}[p]
	\centering
		\subfloat[leading $p_T$]{\includegraphics[width=0.49\textwidth]{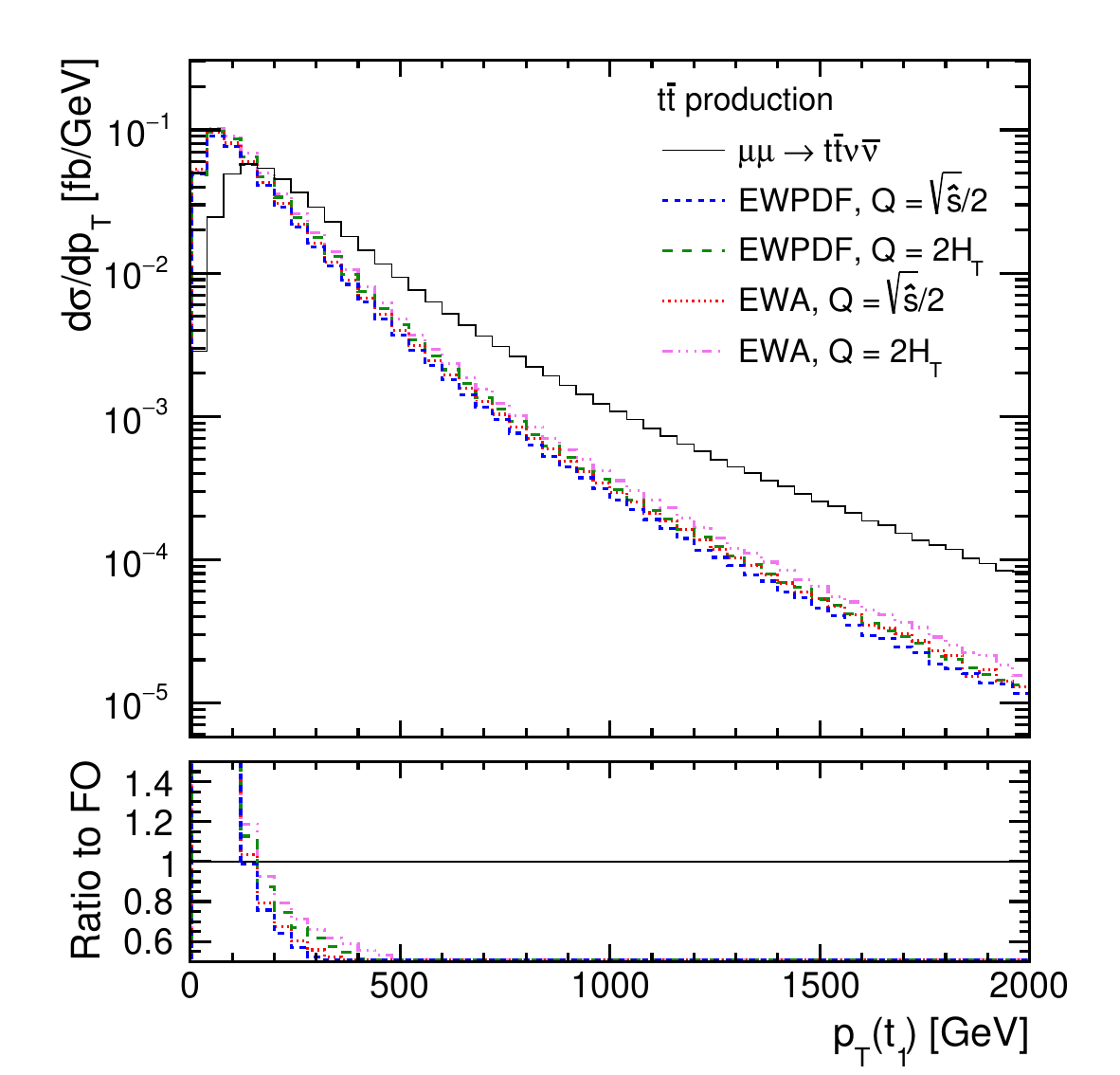}}
		\subfloat[subleading $p_T$]{\includegraphics[width=0.49\textwidth]{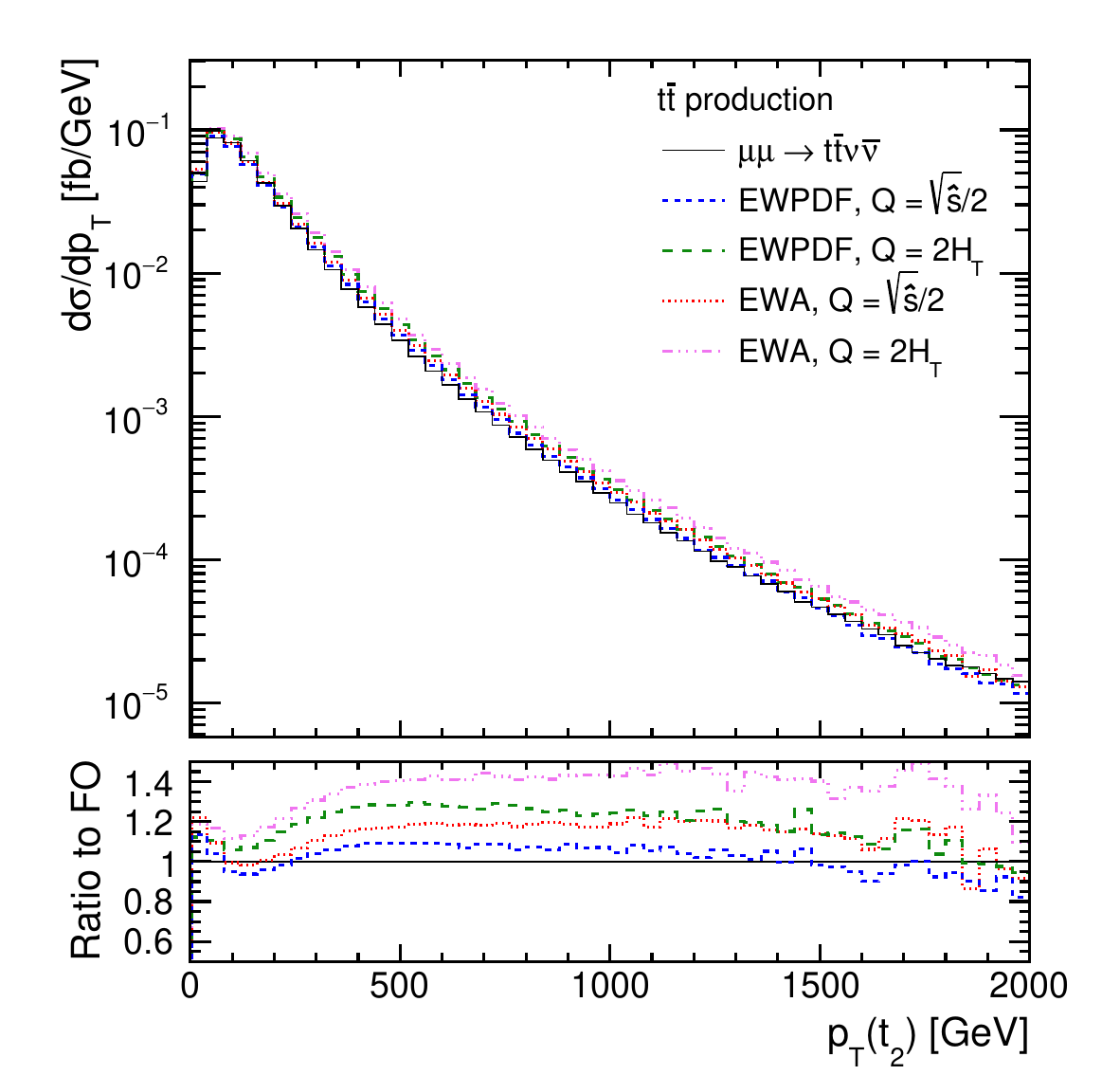}}
    \hfill
		\subfloat[leading $p_T$, $m_{t\bar{t}} >2$\,TeV]{\includegraphics[width=0.49\textwidth]{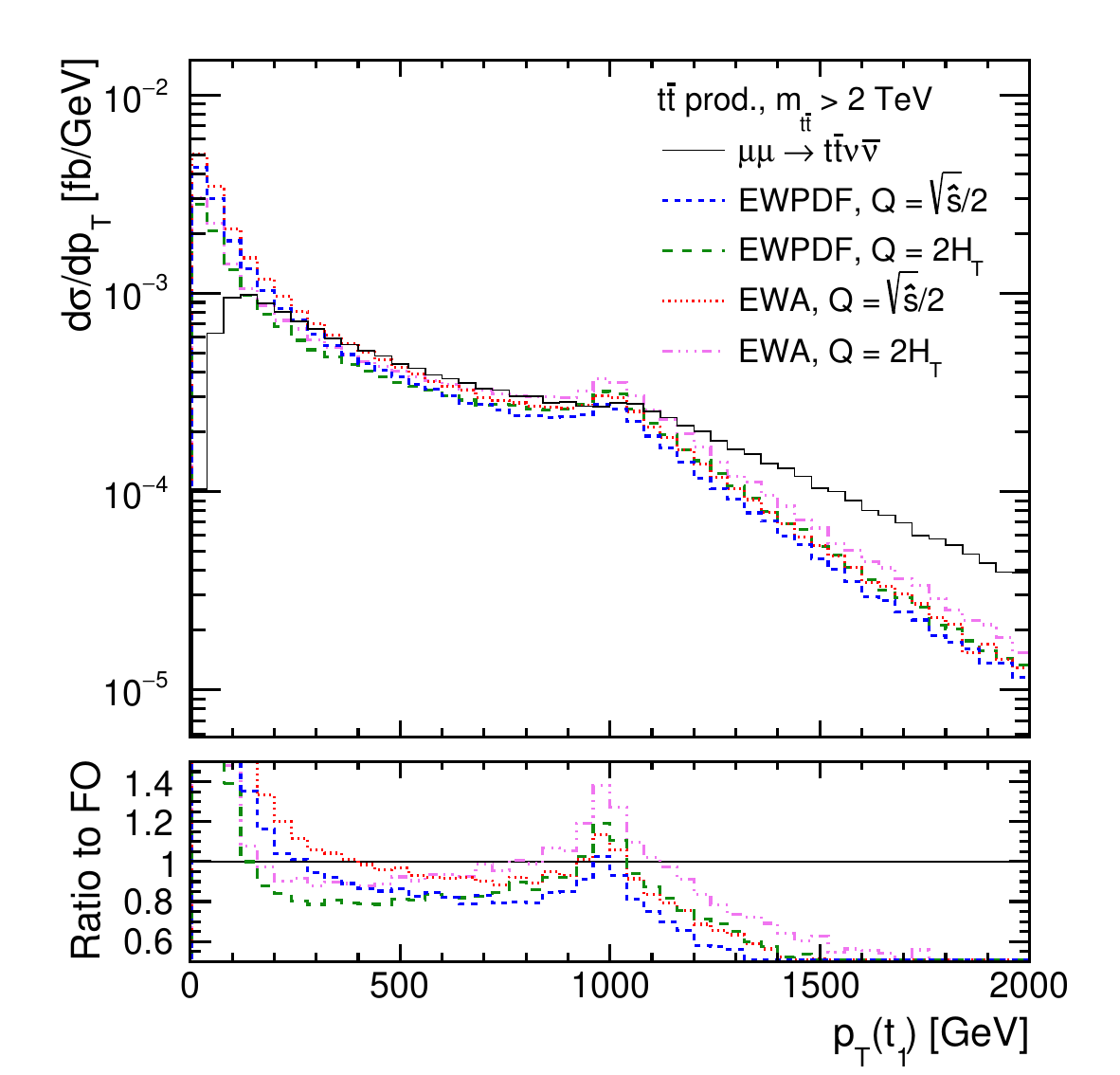}}
		\subfloat[subleading $p_T$, $m_{t\bar{t}} >2$\,TeV]{\includegraphics[width=0.49\textwidth]{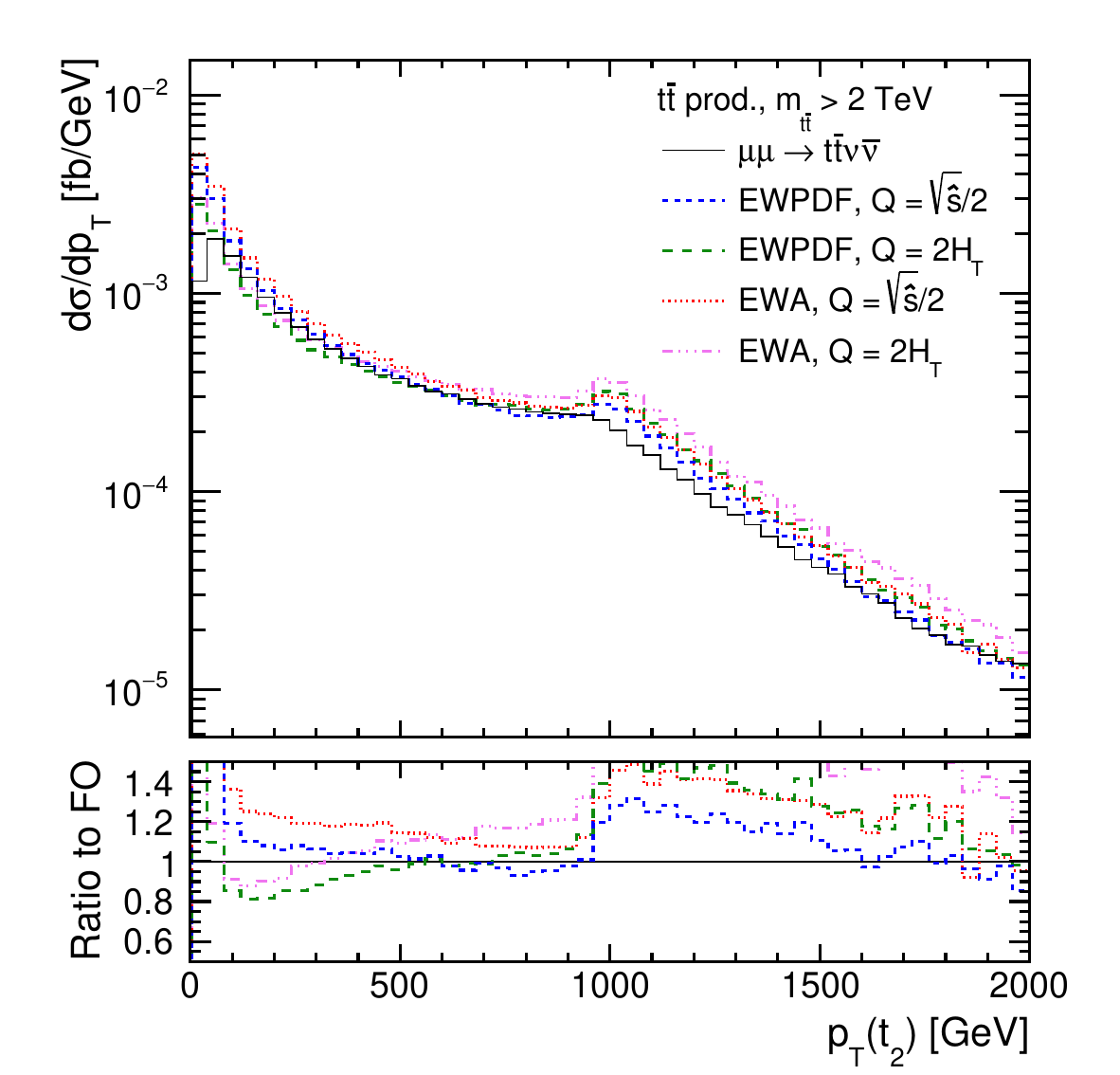}}
	\caption{Transverse momentum distributions of the leading (left panels) and subleading (right panels) partons in the process $\mu^- \mu^+ \to t\bar{t}\nu_\mu \bar{\nu}_\mu$ at $\sqrt{s} = 10$\,TeV for the full matrix element calculation (black), the EW PDF and the EWA with (lower panels) or without (upper panels) an invariant-mass cut of 2 TeV. The blue curve indicates the EW PDF calculation with a scale of $\sqrt{\hat{s}}/2$ and the green curve with a scale of $2H_{T}$; the red curve stands for the EWA calculation with a scale choice of $\sqrt{\hat{s}}/2$, while the pink one --- a scale of $2H_{T}$. The bottom part of each of the four panels shows the ratio to the fixed-order calculation.}
	\label{fig:EWPDF_ttbar_pT}
\end{figure}

\begin{figure}
	\centering
		\subfloat[$m_{t\bar{t}} < 5$\,TeV, $p_T$ > 100 GeV]{\includegraphics[width=0.49\textwidth]{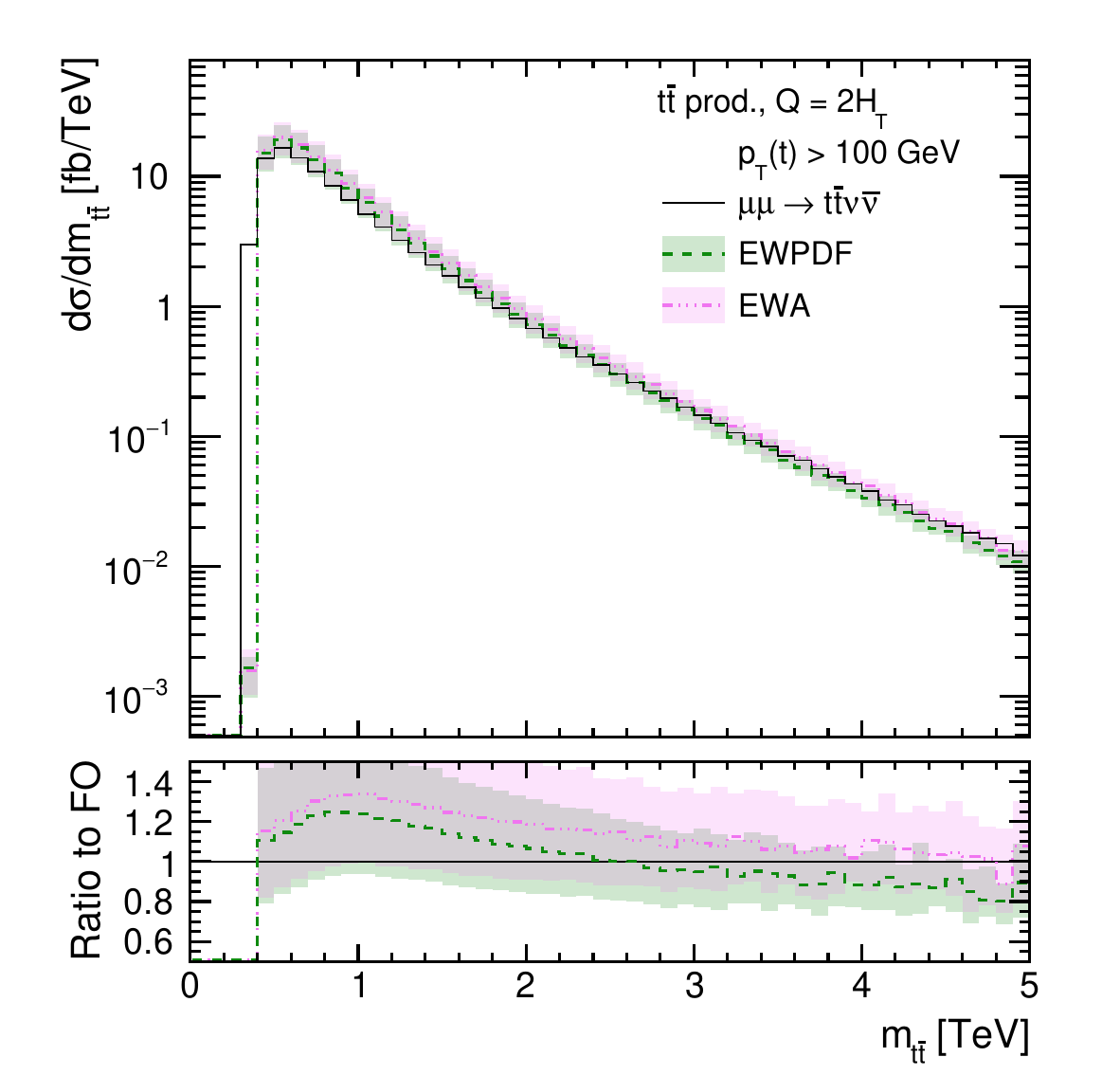}}
	\hfill
		\subfloat[$m_{t\bar{t}} > 5$\,TeV, $p_T$ > 100 GeV]{\includegraphics[width=0.49\textwidth]{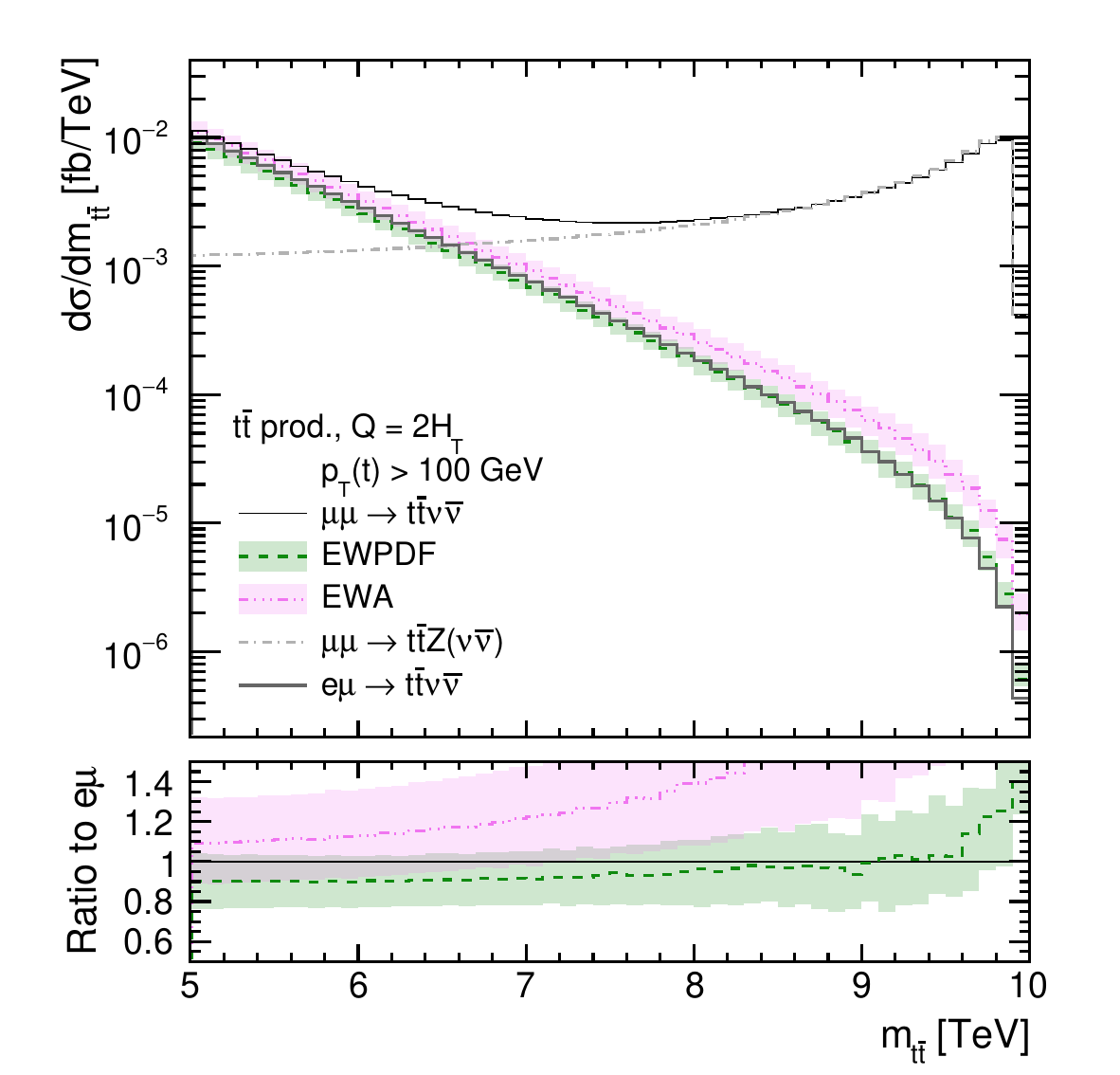}}
	\caption{Invariant mass distributions of the $t \bar{t}$-system at a $\mu^- \mu^+$ collider in the $W^+W^-$ interaction for the full matrix element calculation (black), the EW PDF and the EWA, including different sets of cuts. The green curve shows the EW PDF results with a scale of $2H_T$ and the pink one --- the EWA results with the same scale. The bands represent a variation of the scale by a factor of two. Two additional lines are shown for high invariant masses: dark gray for the full process $e^- \mu^+ \to t\bar{t}\nu_e \bar{\nu}_\mu$ and dash-dotted light gray for the process $\mu^- \mu^+ \to Zt\bar{t}$ with a subsequent decay $Z\to \nu_\mu \bar{\nu}_\mu$. The left bottom panel shows the ratio to the fixed-order calculation, while the right panel shows the ratio to the fixed-order calculation with $e\mu$ beams to reduce non-VBF topologies.}
	\label{fig:EWPDF_ttbar_bands}
\end{figure}

\begin{figure}
	\centering
		\includegraphics[width=0.6\textwidth]{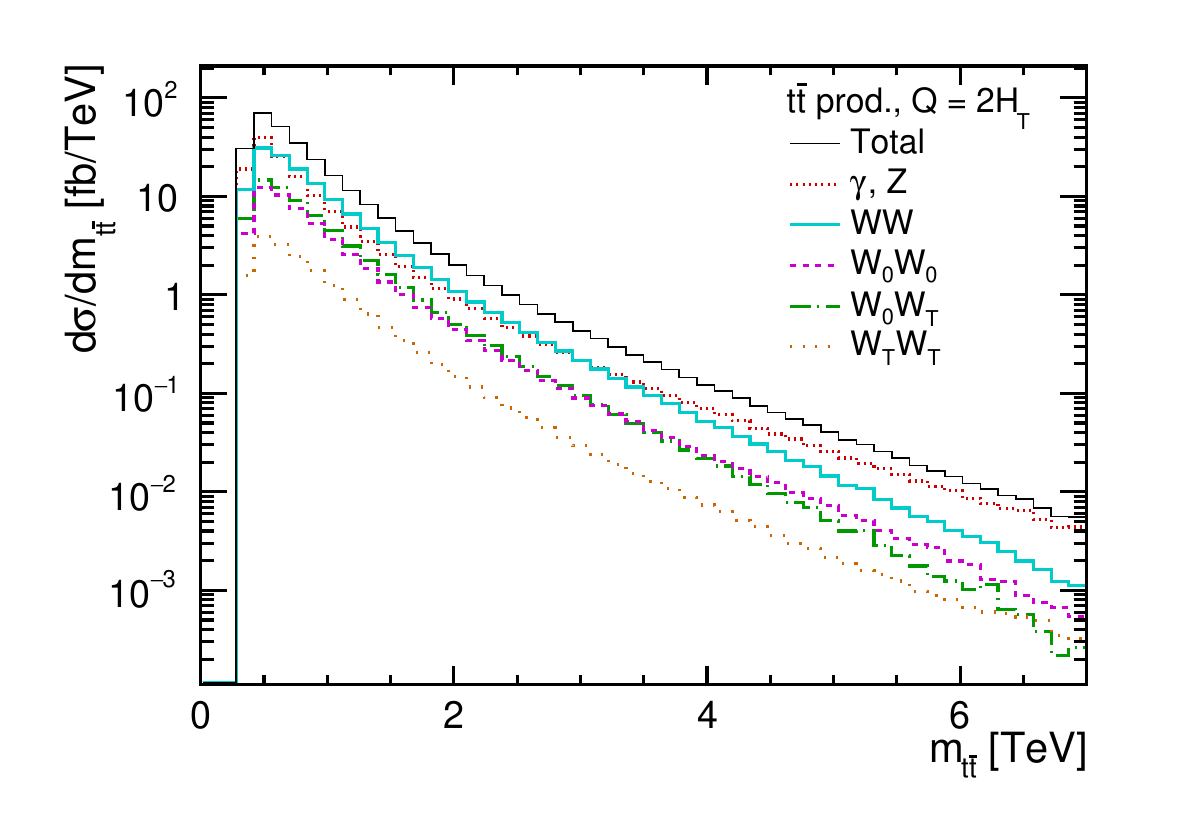}
	\caption{Invariant mass distributions of the $t \bar{t}$-system at a $\mu^- \mu^+$ collider for the sum of processes (black), the neutral-current (red) and the charged-current (cyan) production within the EW PDF framework. The $WW$ process is further separated into the $W_0W_0$ (magenta), $W_0W_T$ (dark green) and $W_TW_T$ (orange) contributions. }
	\label{fig:EWPDF_ttbar_all}
\end{figure}

Next, we study the top-pair production, $\mu^- \mu^+ \to t \bar{t} \nu_\mu\bar{\nu}_\mu$. 
Fig.~\ref{fig:EWPDF_ttbar} shows the top-pair invariant-mass distribution, broken down into the low-energy and high-energy ranges.   We compare two schemes of setting the factorization scale, $\sqrt{\hat{s}}/2$ and $2H_T = 2\sqrt{m_t^2 + p_T^2}$. The latter scheme is inspired by the transverse-mass scale often used at hadron colliders for massive final states with significant transverse momentum. We find that the agreement between the full process and the collinear approximations is reasonably accurate over most of the invariant-mass range.  The factorized approximation yields a rate which is somewhat higher than the full matrix element, but the agreement is satisfactory if we take scale variations into account.  The scheme with the scale choice $2H_T$ appears to be preferred, and it actually matches the $e^-\mu^+$ curve at large invariant masses. As noted previously, the deviations between the full $\mu^-\mu^+$ process and the approximations at high invariant masses are driven by annihilation topologies, which would be absent if an $e^-\mu^+$ collider were considered.  We also observe that the EW PDF results are lower than the corresponding EWA results, as in the $HH$ case discussed above.  However, for a firm conclusion, one would need better control of the uncertainty associated with the scale and scheme choices.

Fig.~\ref{fig:EWPDF_ttbar_pT} shows the transverse momentum distribution for the $t\bar{t}$ production process. The subleading top quark is better described in the collinear approximation. Yet, when applying an invariant-mass cut, the agreement tends to improve for intermediate values of the transverse momentum from about 100 GeV to the typical momentum scale associated with the cut, $M_\text{cut}/2$. As illustrated by Fig.~\ref{fig:EWPDF_ttbar_pT}, while kinematics is not perfectly described within the collinear framework, lower bounds on the transverse momentum should, in general, improve the agreement.  At low $p_T$ we cannot neglect the $W$ mass, hence, the approximation breaks down when looking at differential distributions.\footnote{Another approach involves, for instance, a pseudorapidity cut on the subleading parton~\cite{Dahlen:2025udl}. While applying an experimentally driven cut to an individual particle is rather unconventional, it provides a temporary solution to the simplified kinematics of the EWA and EW PDF frameworks.}

To justify our findings, Fig.~\ref{fig:EWPDF_ttbar_bands} includes the scale uncertainty for the $2H_T$ choice, varied by a factor of two up and down, with a $p_T$ cut of 100 GeV applied to both tops. The scale impact is modest but more pronounced than in the di-Higgs case. The cut shifts the results towards the full calculation, and the agreement remains near perfect in the entire range of invariant masses.

As mentioned earlier, since the beam remnants may escape detection, different initial-state configurations can lead to experimentally indistinguishable final states. Inclusive observables must therefore account for all possible initial states contributing to the given final state. As an example, Fig.~\ref{fig:EWPDF_ttbar_all} shows the contributions of neutral- and charged-current interactions to the $t\bar{t}$ production. The dominant contribution across the entire invariant-mass range arises from $\gamma$ and $Z$ interactions, closely followed by the $WW$ channels. The difference between the two contributions increases with energy. The $WW$ contributions are further separated according to their helicities, showing that the longitudinal-longitudinal and longitudinal-transverse channels are of comparable size. These results are consistent with~\cite{Han:2020uid}.

\subsection{Tau neutrino pair production}
\begin{figure}
	\centering
		\subfloat[$0.5\,\text{TeV}<m_{\nu_\tau\bar{\nu}_\tau} < 5$\,TeV]{\includegraphics[width=0.49\textwidth]{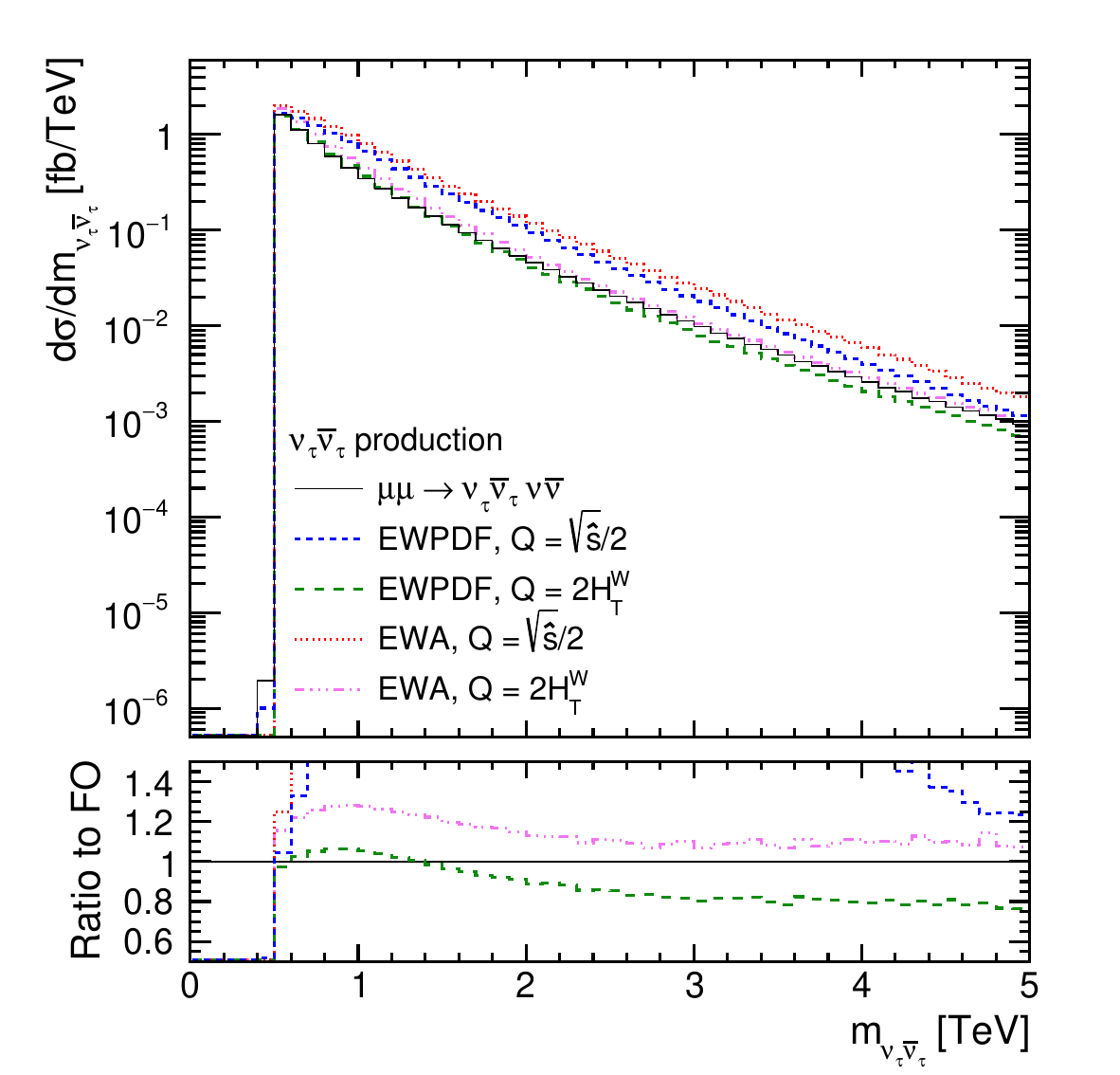}}
	\hfill	\subfloat[$m_{\nu_\tau\bar{\nu}_\tau} > 5$\,TeV]{\includegraphics[width=0.49\textwidth]{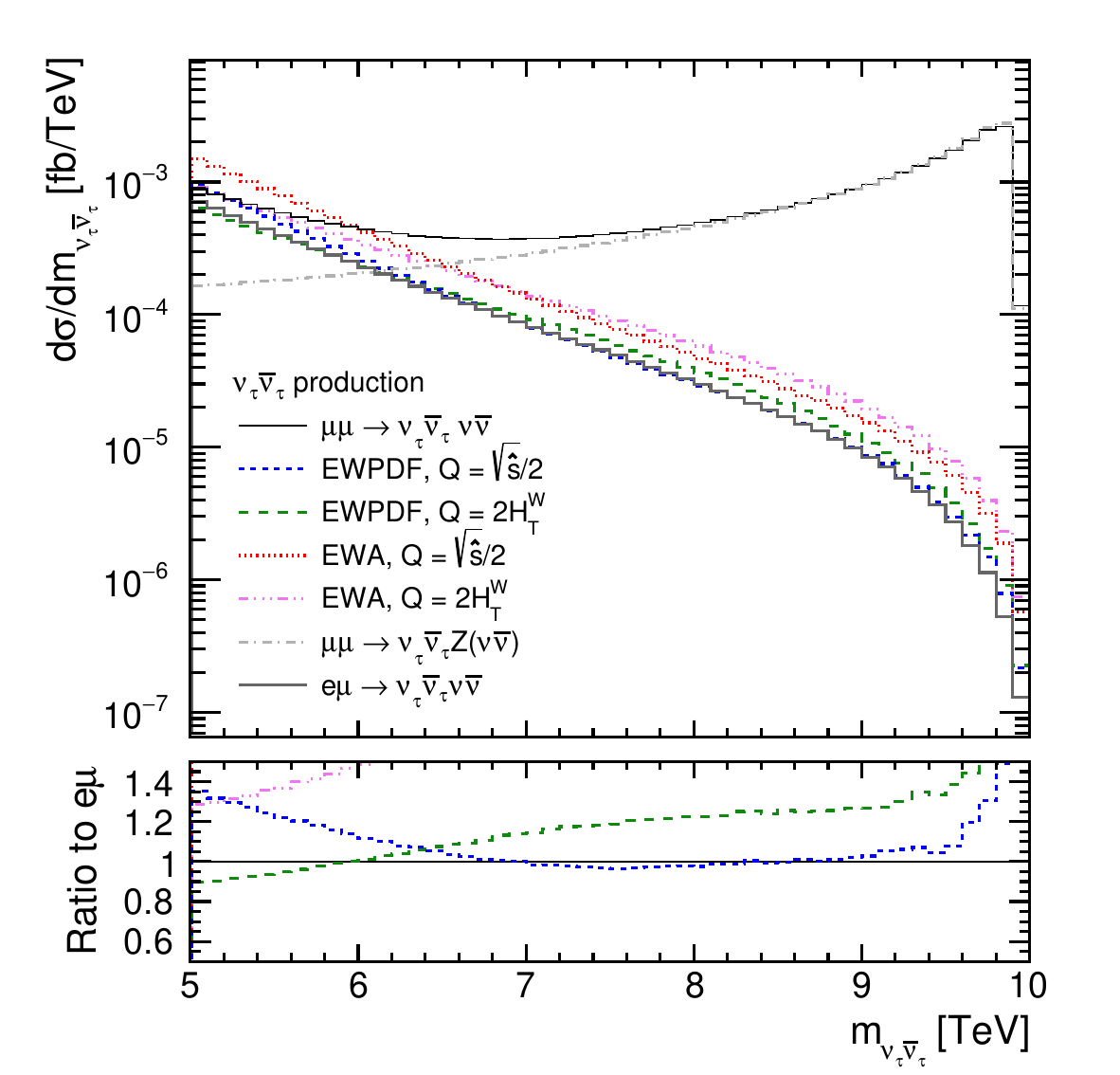}}
	\caption{Invariant mass distributions of the $\nu_\tau \bar{\nu}_\tau$-system at a $\mu^- \mu^+$ collider in the $W^+W^-$ interaction for the full matrix element calculation (black), the EW PDF and the EWA. The blue curve indicates the EW PDF calculation with a scale of $\sqrt{\hat{s}}/2$ and the green curve with a scale of $2H^W_{T}$; the red curve stands for the EWA calculation with a scale choice of $\sqrt{\hat{s}}/2$, while the pink one --- a scale of $2H^W_{T}$. Two additional lines are shown for high invariant masses: dark gray for the full process $e^- \mu^+ \to \nu_\tau \bar{\nu}_\tau\nu_e \bar{\nu}_\mu$ and dash-dotted light gray for the process $\mu^- \mu^+ \to Z\nu_\tau \bar{\nu}_\tau$ with a subsequent decay $Z\to \nu_\mu \bar{\nu}_\mu$. The left bottom panel shows the ratio to the fixed-order calculation, while the right panel shows the ratio to the fixed-order calculation with $e\mu$ beams to reduce non-VBF topologies.}
	\label{fig:EWPDF_vv}
\end{figure}

\begin{figure}
	\centering
		\subfloat[leading $p_T$, $m_{\nu_\tau\bar{\nu}_\tau} >0.5$\,TeV]{\includegraphics[width=0.49\textwidth]{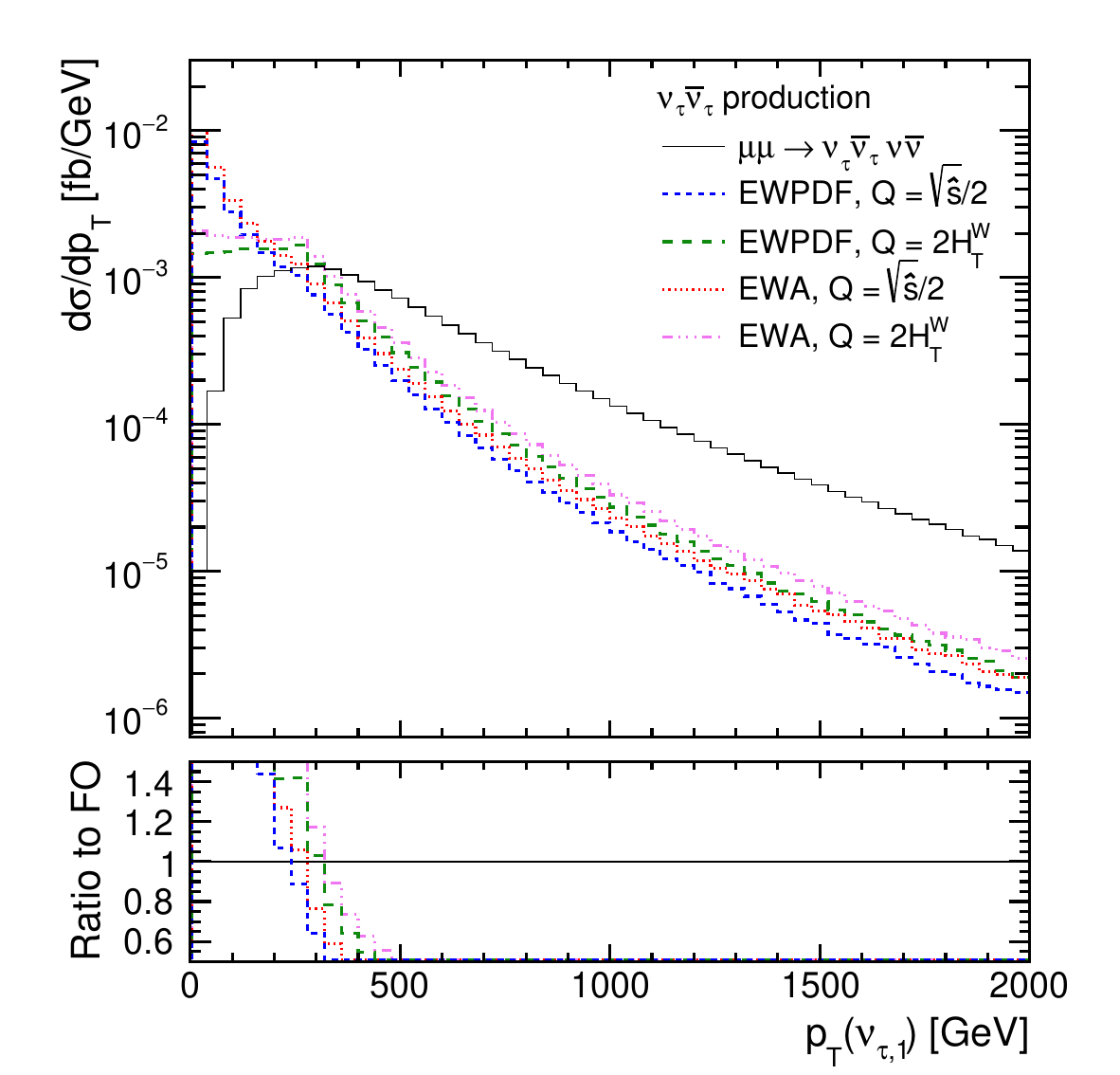}}
		\subfloat[subleading $p_T$, $m_{\nu_\tau\bar{\nu}_\tau} >0.5$\,TeV]{\includegraphics[width=0.49\textwidth]{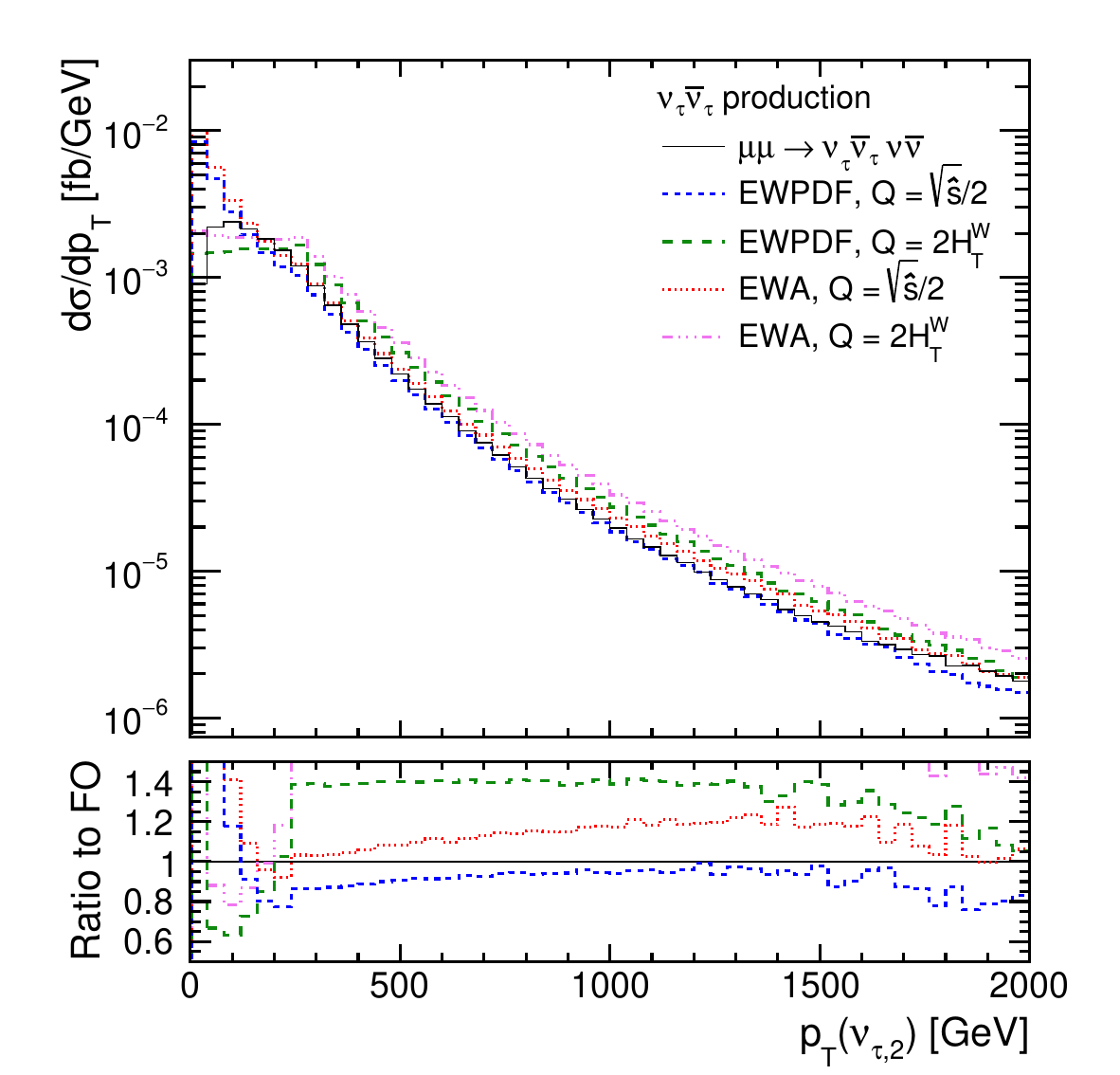}}
    \hfill
		\subfloat[leading $p_T$, $m_{\nu_\tau\bar{\nu}_\tau} >2$\,TeV]{\includegraphics[width=0.49\textwidth]{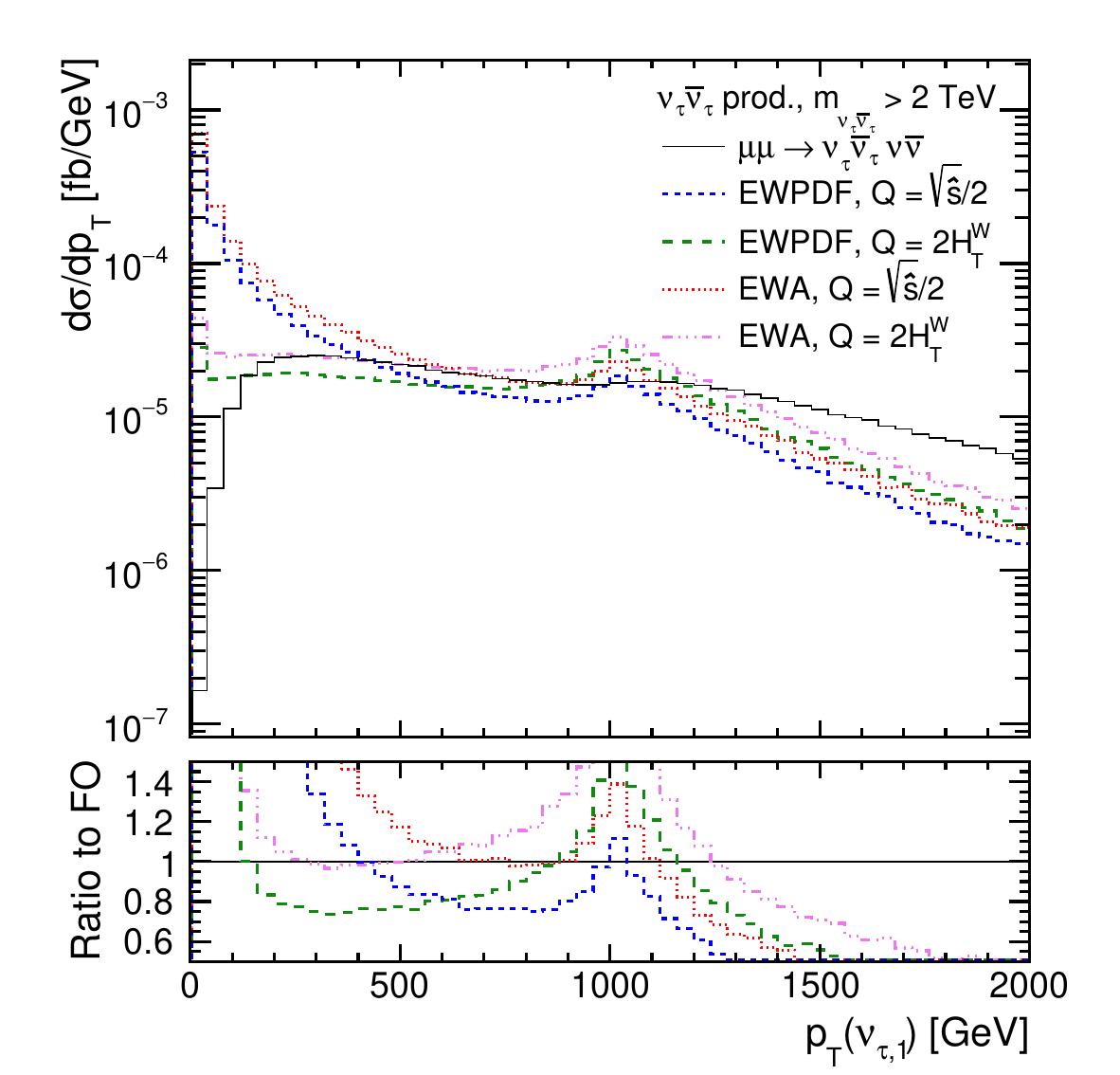}}
	\hfill
		\subfloat[subleading $p_T$, $m_{\nu_\tau\bar{\nu}_\tau} >2$\,TeV]{\includegraphics[width=0.49\textwidth]{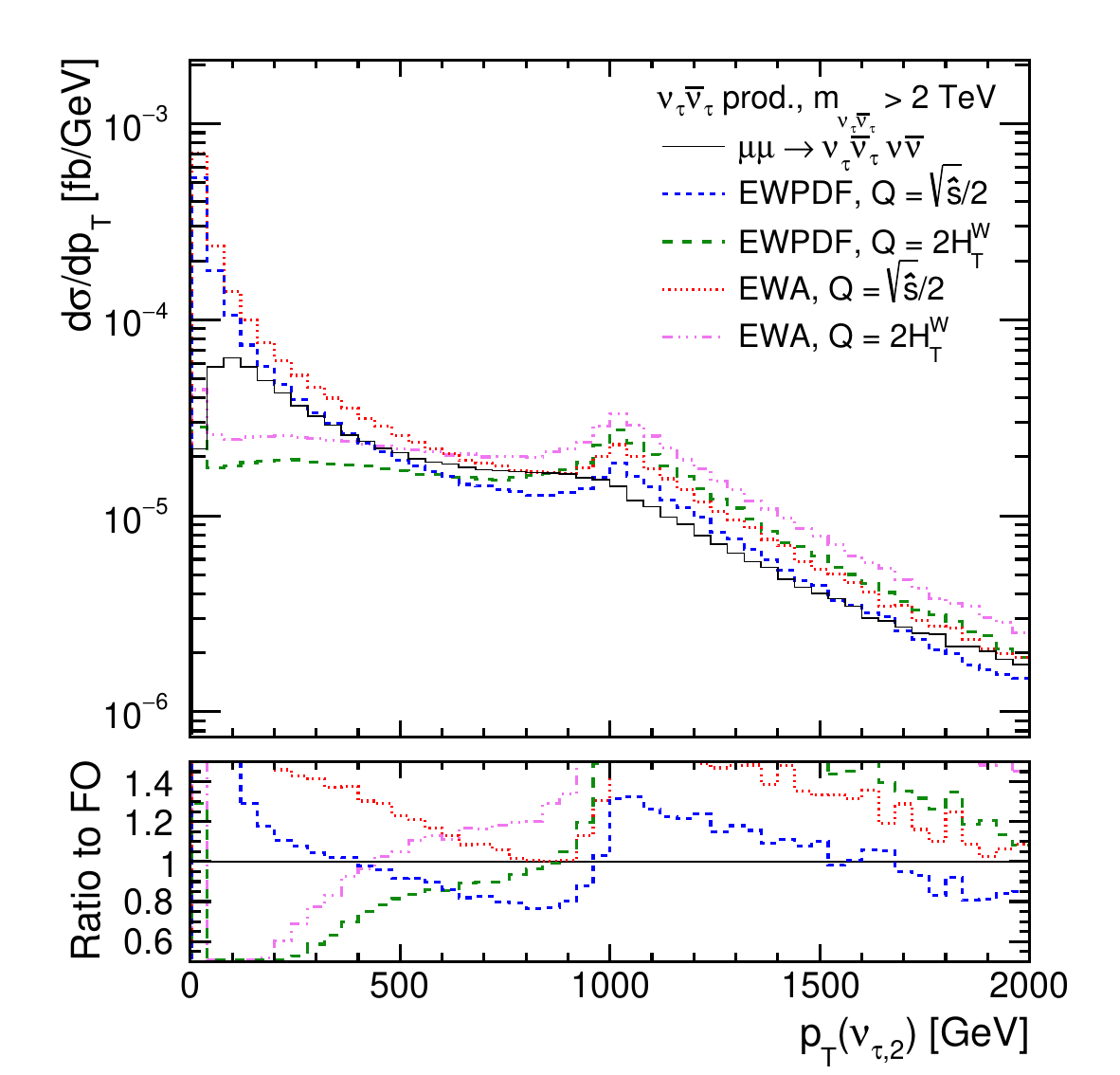}}
	\caption{Transverse momentum distributions of the leading (left panels) and subleading (right panels) partons in the process $\mu^- \mu^+ \to \nu_\tau \bar{\nu}_\tau \nu_\mu \bar{\nu}_\mu$ at $\sqrt{s} = 10$\,TeV for the full matrix element calculation (black), the EW PDF and the EWA with (lower panels) or without (upper panels) an invariant-mass cut of 2 TeV. The blue curve indicates the EW PDF calculation with a scale of $\sqrt{\hat{s}}/2$ and the green curve with a scale of $2H^W_{T}$; the red curve stands for the EWA calculation with a scale choice of $\sqrt{\hat{s}}/2$, while the pink one --- a scale of $2H^W_{T}$. The bottom parts of each panel show the ratio to the fixed-order calculation.}
	\label{fig:EWPDF_vv_pT}
\end{figure}

\begin{figure}
	\centering
		\subfloat[$0.5\,\text{TeV} < m_{\nu_\tau\bar{\nu}_\tau} < 5$\,TeV, $p_T$ > 100 GeV]{\includegraphics[width=0.49\textwidth]{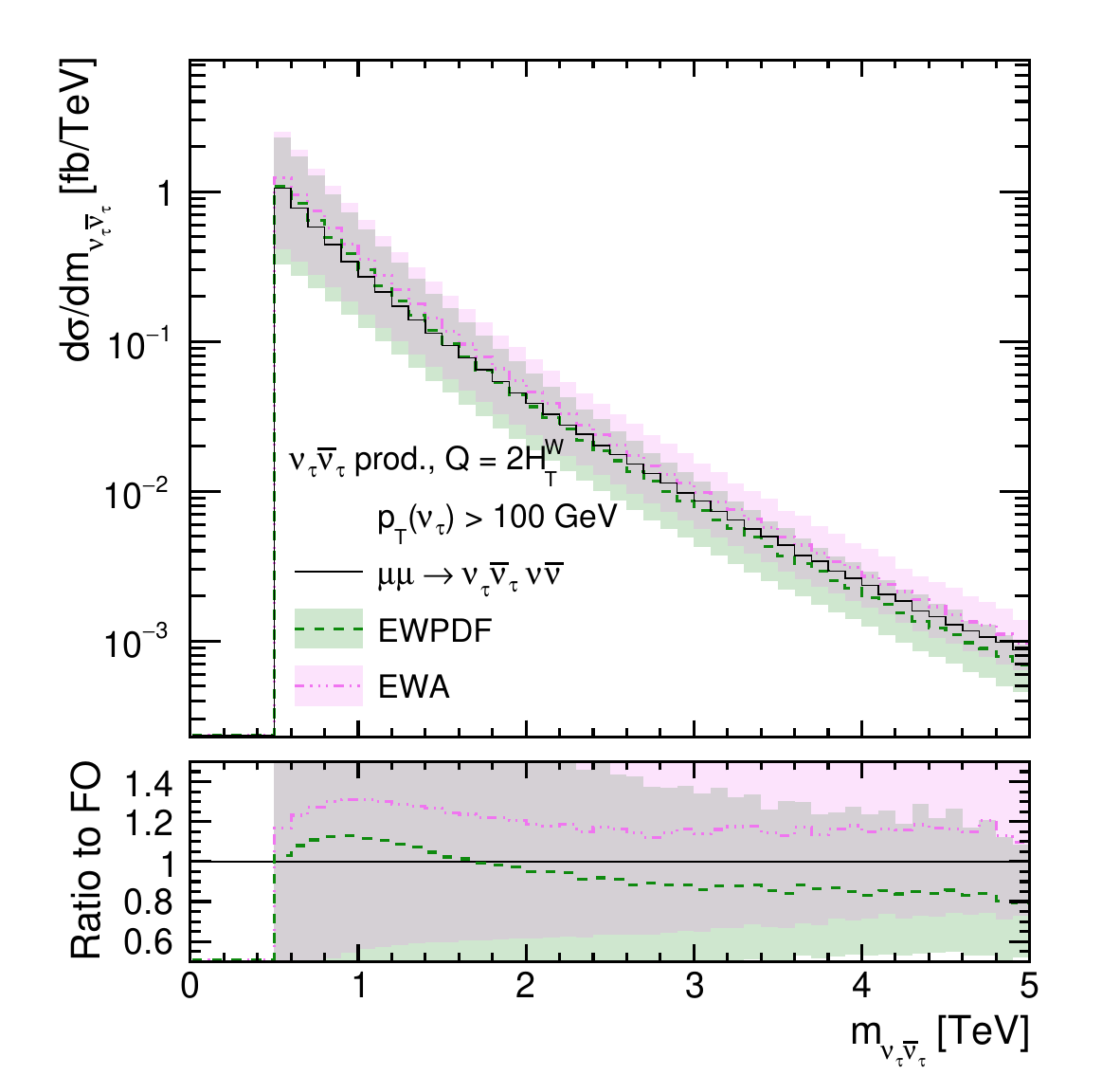}}
	\hfill		\subfloat[$m_{\nu_\tau\bar{\nu}_\tau} > 5$\,TeV, $p_T$ > 100 GeV]{\includegraphics[width=0.49\textwidth]{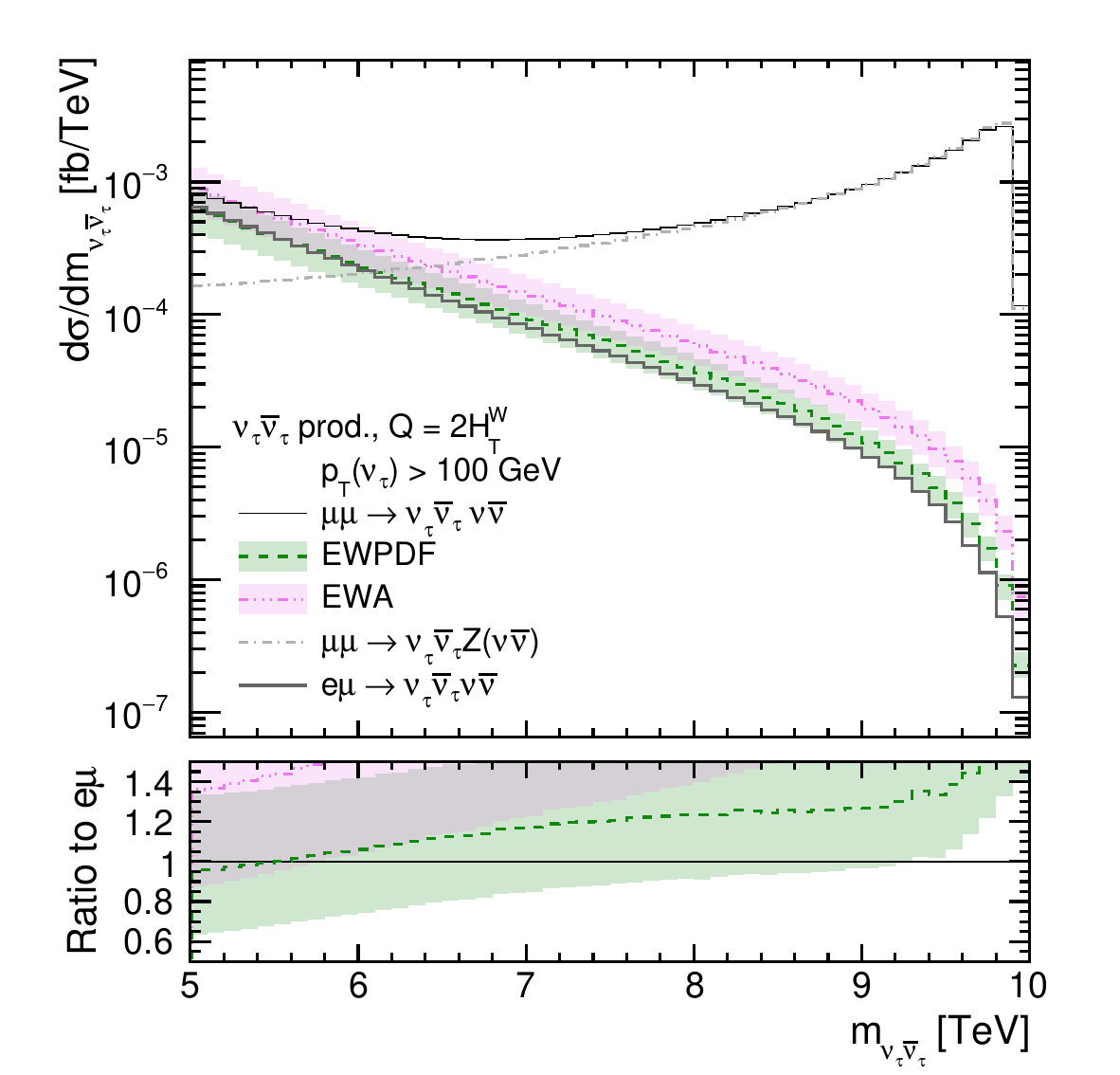}}
	\caption{Invariant mass distributions of the $\nu_\tau \bar{\nu}_\tau$-system at a $\mu^- \mu^+$ collider in the $W^+W^-$ interaction for the full matrix element calculation (black), the EW PDF and the EWA, including different sets of cuts. The green curve shows the EW PDF results with a scale of $2H^W_T$ and the pink one --- the EWA results with the same scale. The bands represent a variation of the scale by a factor of two. Two additional lines are shown for high invariant masses: dark gray for the full process $e^- \mu^+ \to \nu_\tau \bar{\nu}_\tau\nu_e \bar{\nu}_\mu$ and dash-dotted light gray for the process $\mu^- \mu^+ \to Z\nu_\tau \bar{\nu}_\tau$ with a subsequent decay $Z\to \nu_\mu \bar{\nu}_\mu$. The left bottom panel shows the ratio to the fixed-order calculation, while the right panel shows the ratio to the fixed-order calculation with $e\mu$ beams to reduce non-VBF topologies.}
	\label{fig:EWPDF_vv_bands}
\end{figure}

In the next step, we consider the process $\mu^- \mu^+ \to \nu_\tau \bar \nu_\tau \nu_\mu \bar \nu_\mu$, with the tau neutrinos selected to suppress non-VBF topologies. Although not experimentally accessible, the process is dominated by transversely polarized vector bosons and is therefore of theoretical interest. Nevertheless, neutrinos produced via $Z$ boson radiation are always present and cannot be easily removed to separate the VBF-like topologies without violating gauge invariance; their impact is instead mitigated by a lower invariant-mass cut of 500\,GeV applied to the tau-neutrino pair. 

Fig.~\ref{fig:EWPDF_vv} shows the invariant-mass distribution of the tau-neutrino system with two scales, $\sqrt{\hat{s}}/2$ and $2H^W_T = 2\sqrt{m_W^2 + p_T^2}$ at low and high invariant masses\footnote{The $H_T$ variable is modified to account for the smallness of the neutrino masses, which would otherwise lead to an unphysically low scale at vanishing $p_T$. This particular choice reflects the presence of a virtual electroweak boson exchange that mediates the interaction.}. At low invariant masses, the $\sqrt{\hat{s}}/2$ scale seems to perform much worse than the $2H_T$ scale, which remains robust over the entire range of invariant masses. Interestingly, in the high-invariant-mass regime, the $\sqrt{\hat{s}}/2$ scale yields slightly better agreement, closely reproducing the full calculation, yet within the uncertainty bands (see Fig.~\ref{fig:EWPDF_vv_bands}). The overall agreement is remarkably good, demonstrating that the collinear framework can approximately describe the full process. As observed in the other examples, the EW PDF framework appears to outperform the EWA implementation, which likely originates from its superior kinematic treatment of the splitting, as explained in the di-Higgs case.

To compare the two scale choices, we show the transverse-momentum distributions in Fig.~\ref{fig:EWPDF_vv_pT}. The subleading parton is again described more accurately, and the $2H_T^W$ curves are clearly affected by the low invariant-mass cutoff of 500\,GeV. When applying a high invariant-mass cut, the agreement for the subleading parton mildly deteriorates, yet the behavior of the leading parton approaches the full-calculation result. This again motivates a minimal transverse momentum of 100 GeV, as applied in Fig.~\ref{fig:EWPDF_vv_bands}. The figure includes uncertainty bands, which are visibly larger than previously. Yet, the central value of the EW PDF approach closely reproduces the full calculation.

\subsection{ $W$ boson pair production}
\begin{figure}
	\centering
		\subfloat[invariant mass]{\includegraphics[width=0.49\textwidth]{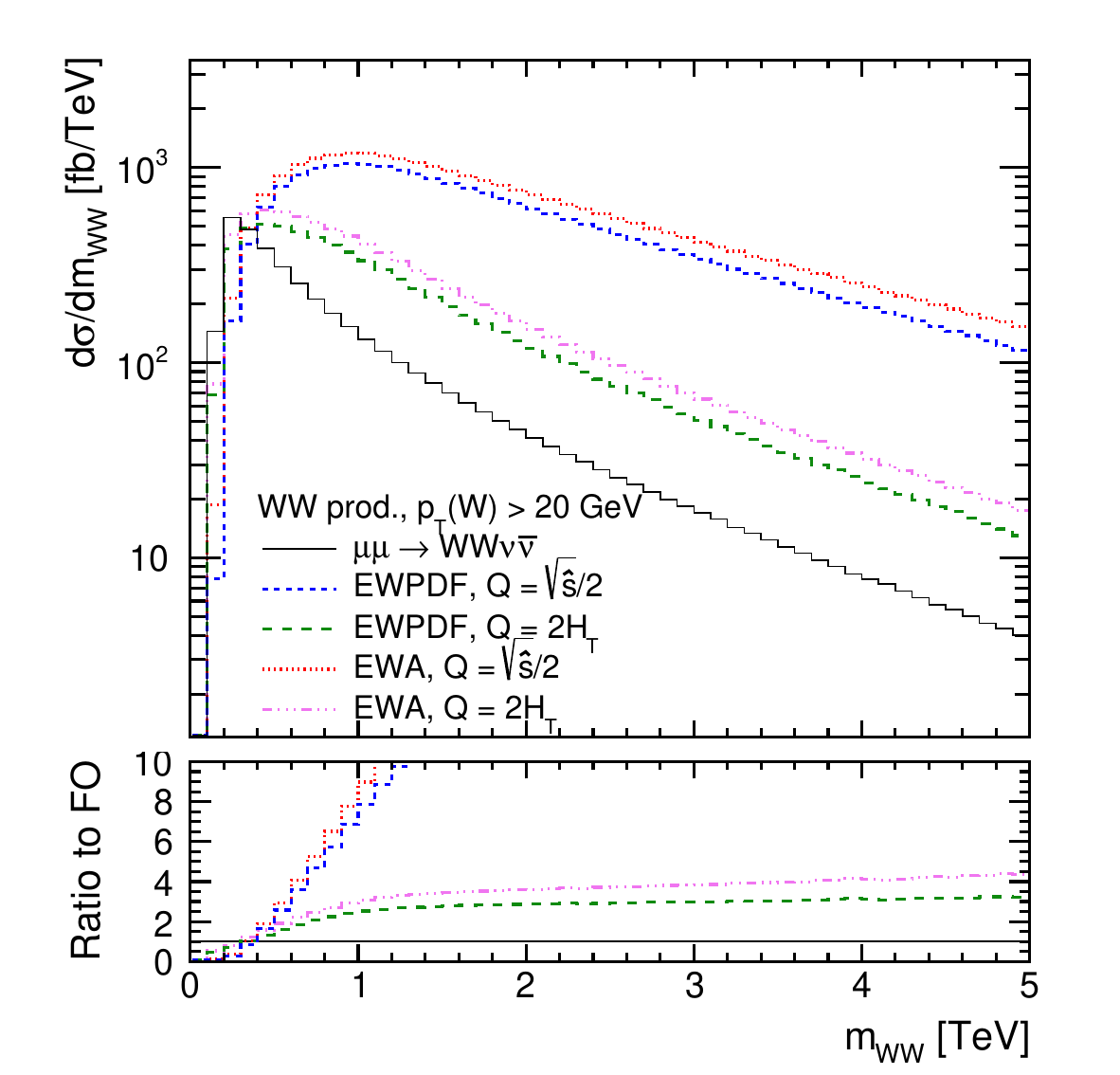}}
	\hfill
		\subfloat[pseudorapidity]{\includegraphics[width=0.49\textwidth]{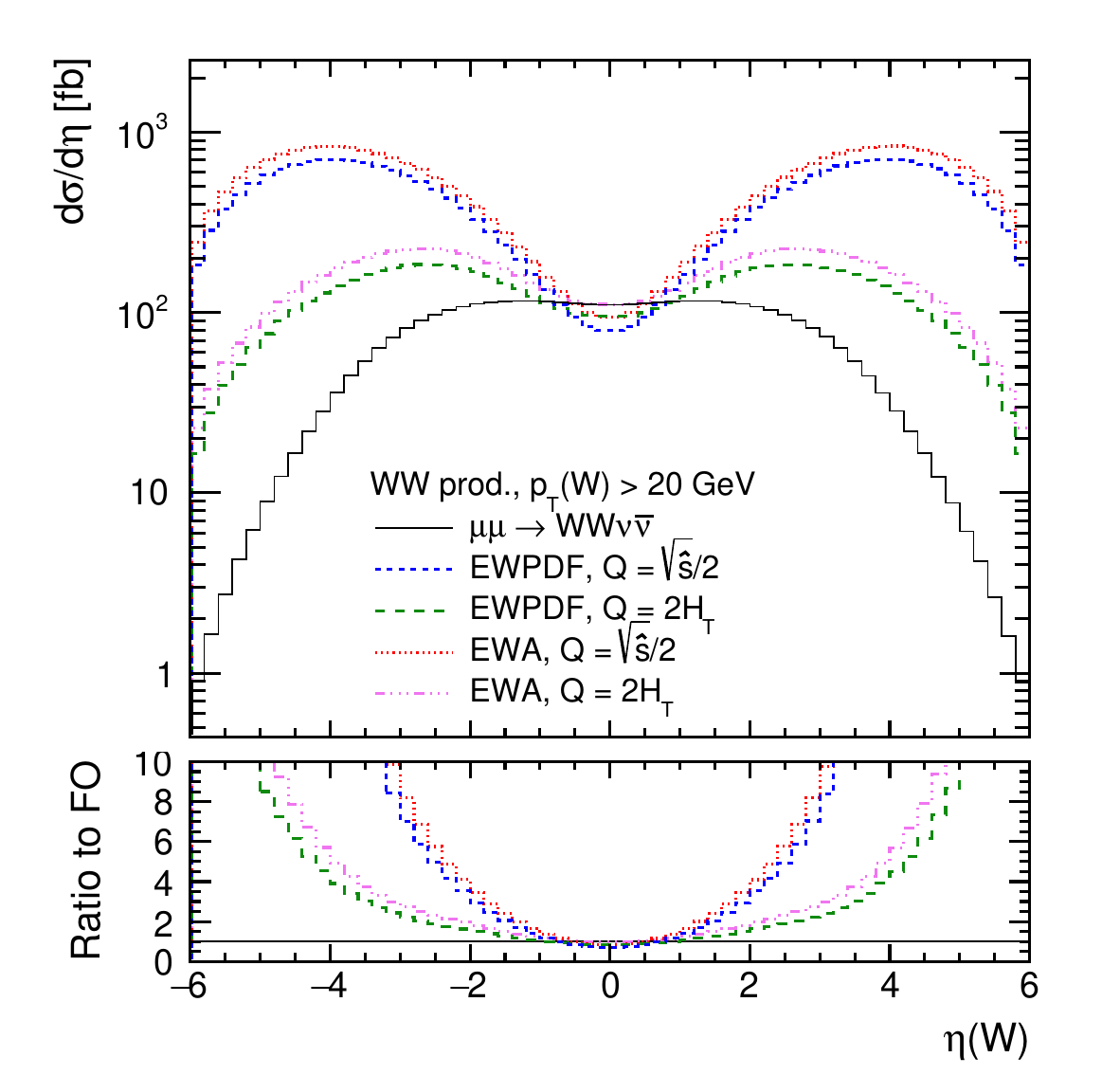}}
	\caption{Left: Invariant mass distributions of the $W^+W^-$ system in the vector boson scattering $\mu^- \mu^+ \to W^+ W^- \nu_\mu \bar{\nu}_\mu$ at $\sqrt{s} = 10$\,TeV. Right: Pseudorapidity distribution for final-state partons in the same process. The full matrix element calculation is shown in black, the blue curve indicates the EW PDF calculation with a scale of $\sqrt{\hat{s}}/2$, and the green curve with a scale of $2H_{T}$; the red curve stands for the EWA calculation with a scale choice of $\sqrt{\hat{s}}/2$, while the pink one --- a scale of $2H_{T}$. The bottom panels show the ratio to the fixed-order calculation.}
	\label{fig:EWPDF_WW}
\end{figure}

\begin{figure}
	\centering
		\subfloat[leading $p_T$]{\includegraphics[width=0.49\textwidth]{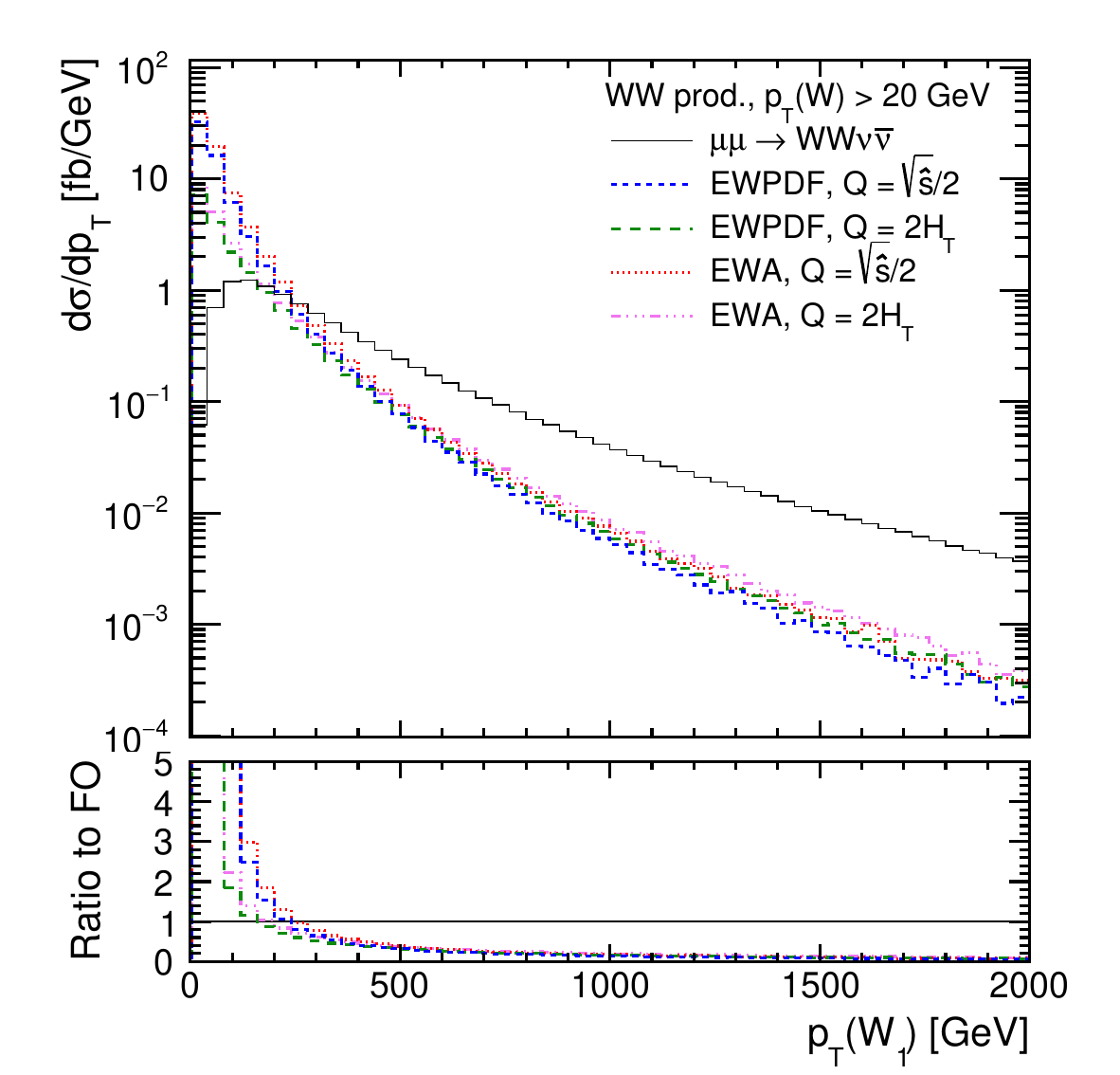}}
	\hfill
		\subfloat[subleading $p_T$]{\includegraphics[width=0.49\textwidth]{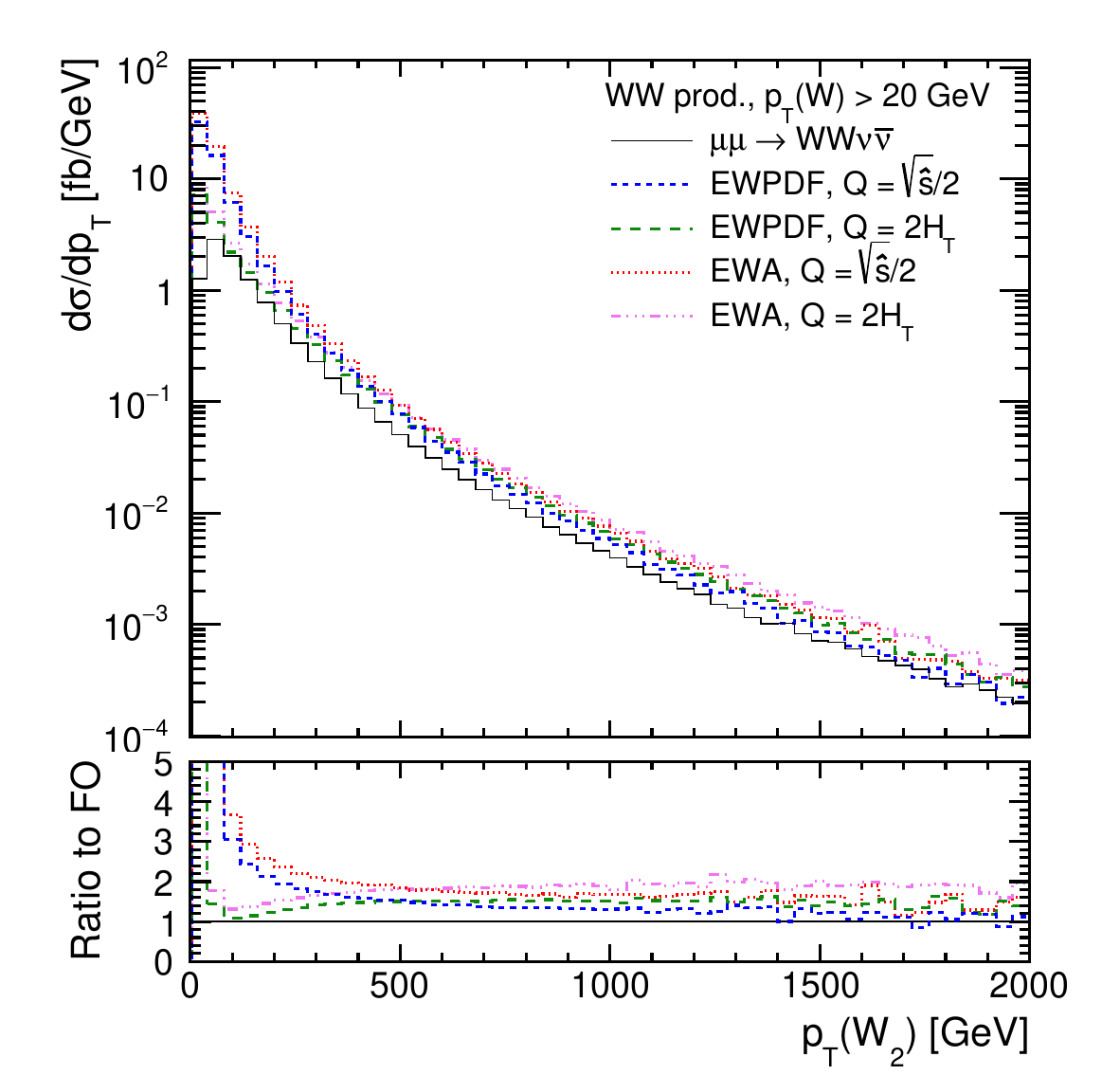}}
	\caption{Transverse momentum distributions of the leading (left panel) and subleading (right panel) partons in the process $\mu^- \mu^+ \to WW\nu_\mu \bar{\nu}_\mu$ at $\sqrt{s} = 10$\,TeV for the full matrix element calculation (black), the EW PDF and the EWA. The blue curve indicates the EW PDF calculation with a scale of $\sqrt{\hat{s}}/2$ and the green curve with a scale of $2H_{T}$; the red curve stands for the EWA calculation with a scale choice of $\sqrt{\hat{s}}/2$, while the pink one --- a scale of $2H_{T}$. The bottom panels show the ratio to the fixed-order calculation.}
	\label{fig:EWPDF_WW_pT}
\end{figure}

\begin{figure}
	\centering
		\subfloat[$m_{WW} < 5$\,TeV, $|\eta| < 2.44$]{\includegraphics[width=0.49\textwidth]{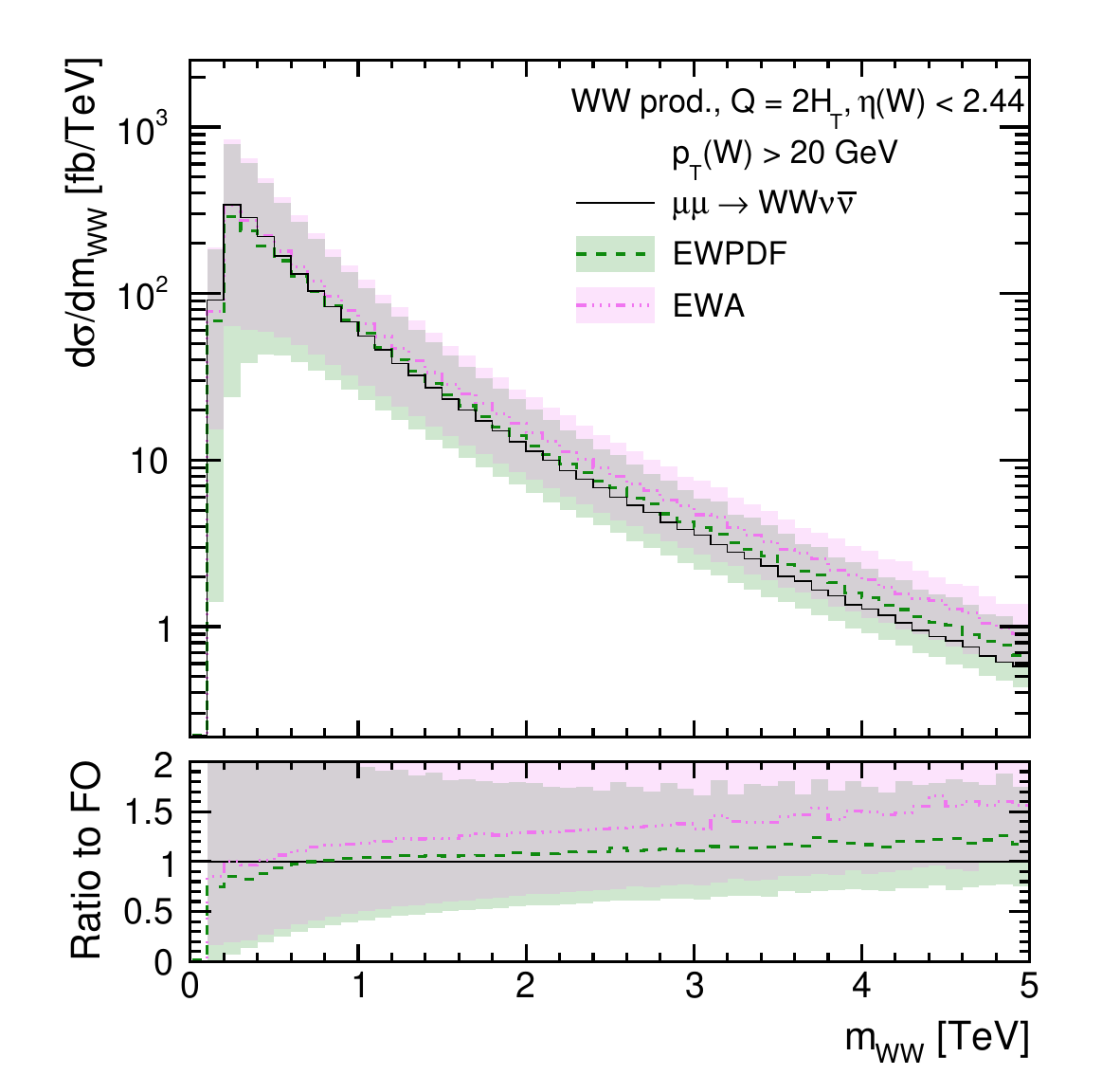}}
	\hfill
		\subfloat[$m_{WW} > 5$\,TeV, $|\eta| < 2.44$]{\includegraphics[width=0.49\textwidth]{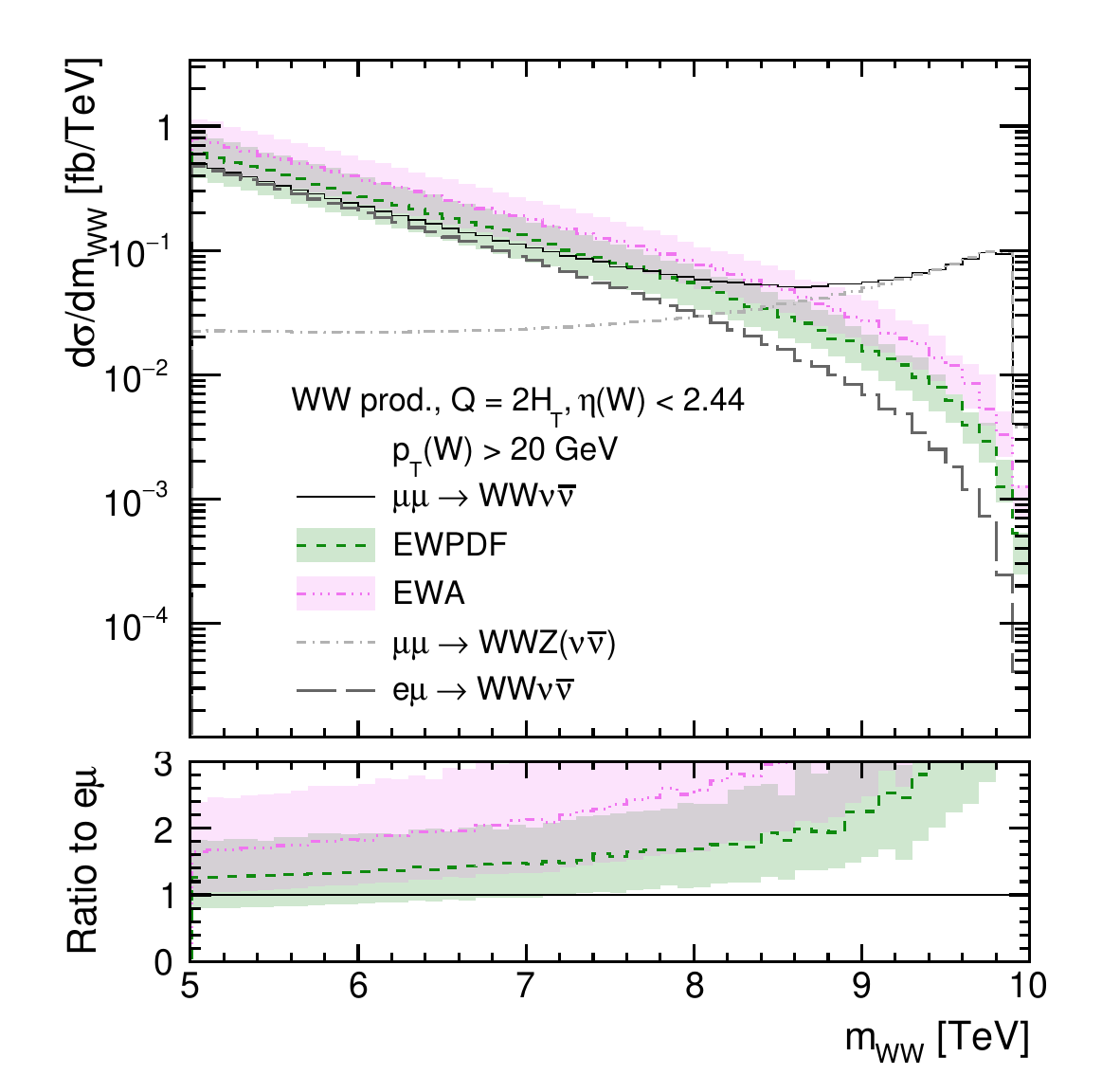}}
	\caption{Invariant mass distributions of the $W^+W^-$ system at a $\mu^- \mu^+$ collider in the process $\mu^- \mu^+ \to WW\nu_\mu \bar{\nu}_\mu$ for the full matrix element calculation (black), the EW PDF and the EWA, including different sets of cuts. The green curve shows the EW PDF results with a scale of $2H_T$ and the pink one --- the EWA results with the same scale. The bands represent a variation of the scale by a factor of two. Two additional lines are shown for high invariant masses: dark gray for the full process $e^- \mu^+ \to WW\nu_e \bar{\nu}_\mu$ and dash-dotted light gray for the process $\mu^- \mu^+ \to ZWW$ with a subsequent decay $Z\to \nu_\mu \bar{\nu}_\mu$. The left bottom panel shows the ratio to the fixed-order calculation, while the right panel shows the ratio to the fixed-order calculation with $e\mu$ beams to reduce non-VBF topologies.}
	\label{fig:EWPDF_WW_cut}
\end{figure}

As a particularly difficult example, we investigate factorization for quasi-elastic vector boson scattering, $\mu^+\mu^- \rightarrow W^+ W^-\nu_\mu \bar \nu_\mu$.  The on-shell subprocess $W^+W^-\to W^+W^-$ has a genuine Coulomb singularity, hence, its total cross section is infinite.  In the full process, the singularity is cut off by the off-shellness of the intermediate $W$ bosons, and the total cross section is finite.  Furthermore, the $W$ boson is a gauge boson of the electroweak theory.  If we restrict the contributing graphs to VBF topologies and project the hard process on-shell, we violate gauge invariance and thus neglect potentially large gauge cancellations as well as topologies which supply positive contributions.  This issue was studied in earlier calculations of strong $W$ scattering, cf.~\cite{Gunion:1986gm, Boos:1999kj, Accomando:2006mc}, where the conclusion was that a matrix-element level calculation of the full process is essential for obtaining reliable results.

In this work, we reconsider this process within the SM with the physical Higgs boson included, comparing the full matrix-element calculation with both the EWA and the EW PDF approximations.  To remove the would-be Coulomb singularity from the approximate result, we apply a generator-level cut of 20\,GeV on the $p_T$ of the final-state $W$ bosons.  In Fig.~\ref{fig:EWPDF_WW} (left), we show the invariant-mass distribution of the $W$-boson pairs. Without additional cuts, the collinear framework overestimates the result by more than an order of magnitude if we choose $\sqrt{\hat s}/2$ as the factorization scale. The $2H_T$ scheme performs noticeably better, but is still above the fixed-order result.  We conclude that the problem of missing gauge cancellations is not eliminated by a $p_T$ cut.  This fact becomes even more evident if we look at kinematics in more detail.  The right panel of Fig.~\ref{fig:EWPDF_WW} displays the pseudorapidity distribution.  The forward/backward region is largely overpopulated, while there are events missing in the central region.

Fig.~\ref{fig:EWPDF_WW_pT} shows the transverse momentum distribution for the leading and subleading partons.  For the leading parton, the approximate results show an enhancement in the low-$p_T$ region which is unphysical, while the high-$p_T$ range is significantly underestimated.  The difference between the two scale-setting schemes appears mostly at low $p_T$.  Incidentally, for the subleading parton (right panel) the description at high $p_T$ is rather accurate and comparable to the processes considered above. 

To further reduce the critical phase-space region, we introduce a pseudorapidity cut, $|\eta| < 2.44$.  The cut corresponds to an angle of 10$^\circ$, and it is well motivated by the current design of the muon collider detectors.  The invariant mass distribution is shown again in Fig.~\ref{fig:EWPDF_WW_cut}. The agreement is very good for low invariant masses, well within the uncertainty bands. However, at high invariant masses, the collinear approximation slowly loses its predictive power. It should also be noted that while the applied cuts are experimentally motivated, the excellent agreement at low invariant masses is somewhat accidental and remains sensitive to the choice of cut values. Nonetheless, once the problematic low-angle region is excluded, the approximation predicts at least the correct order of magnitude. Given the delicate gauge cancellations and complex electroweak structure of this process, the approximation performs relatively well also in this case.

\subsection{Heavy scalar pair production}

\begin{figure}[h]
	\centering
		\subfloat[$m_{XX} < 7$\,TeV]{\includegraphics[width=0.49\textwidth]{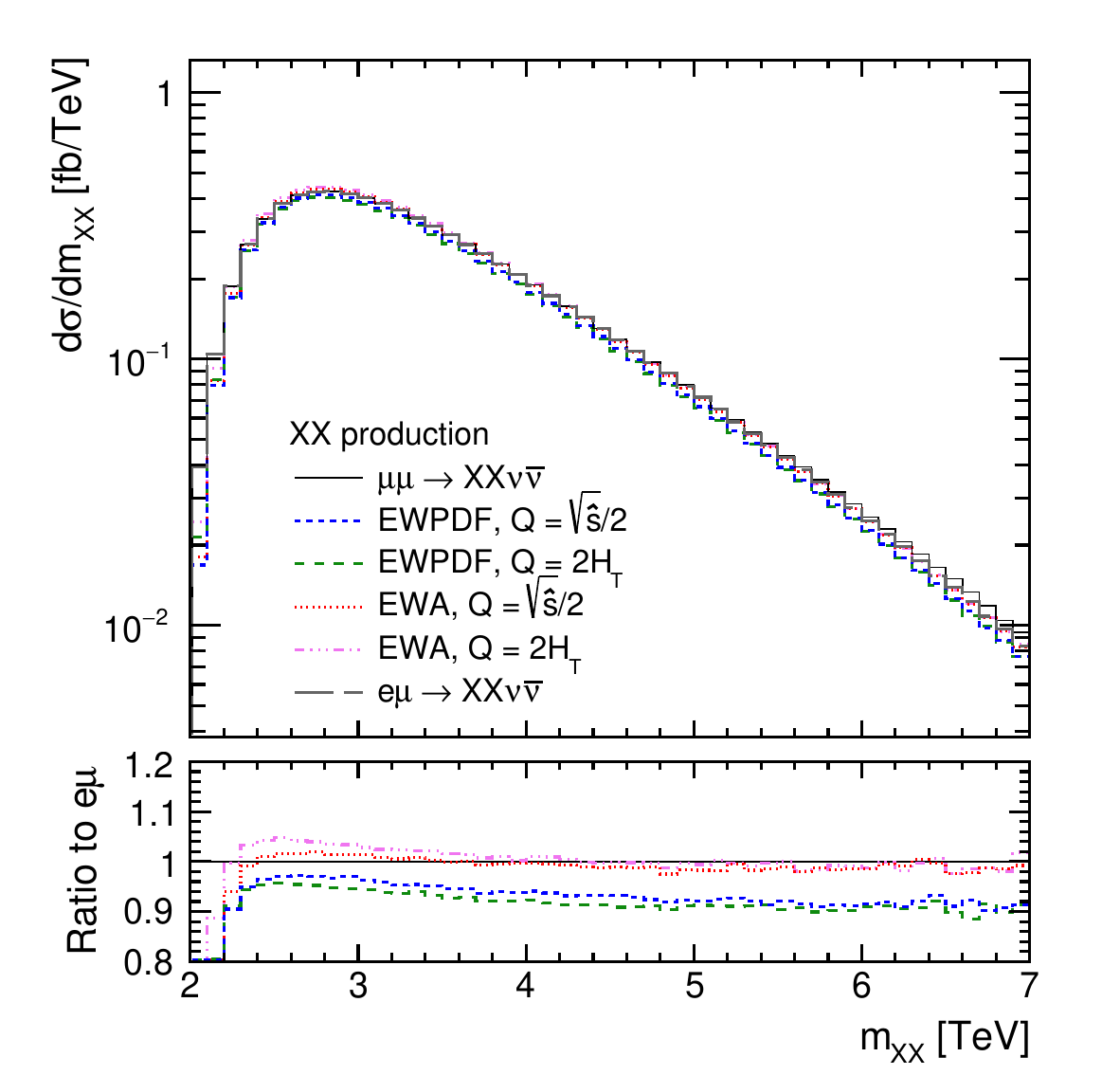}}
	\hfill
		\subfloat[$m_{XX} > 7$\,TeV]{\includegraphics[width=0.49\textwidth]{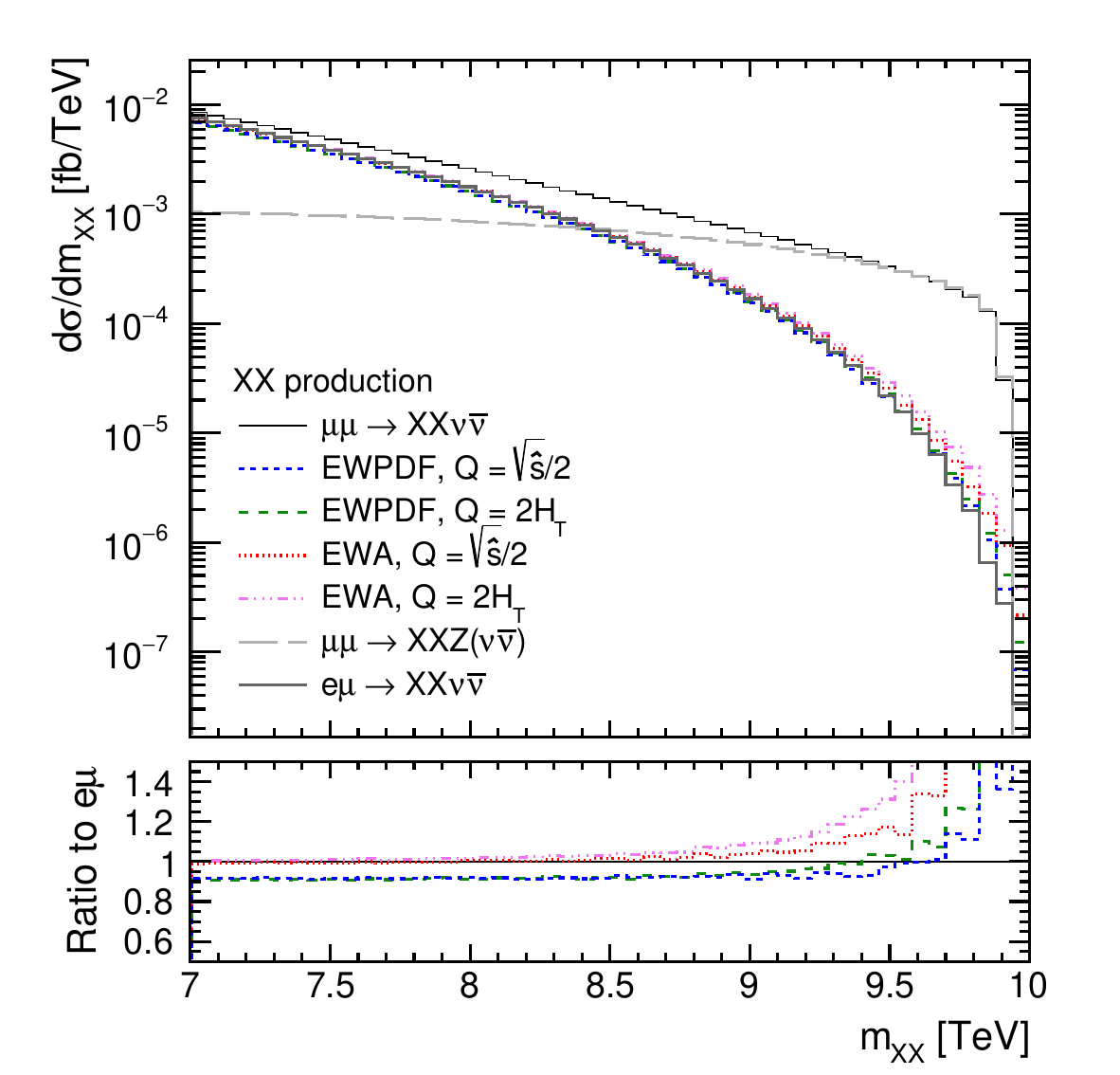}}
	\caption{Invariant mass distributions of the $XX$-system at a $\mu^- \mu^+$ collider in the $W^+W^-$ interaction for the full matrix element calculation (black), the EW PDF and the EWA. The blue curve indicates the EW PDF calculation with a scale of $\sqrt{\hat{s}}/2$ and the green curve with a scale of $2H_{T}$; the red curve stands for the EWA calculation with a scale choice of $\sqrt{\hat{s}}/2$, while the pink one --- a scale of $2H_{T}$. The bands represent a variation of the $\sqrt{\hat{s}}/2$ scale by a factor of two. Two additional lines are shown for high invariant masses: solid dark gray for the full process $e^- \mu^+ \to XX\nu_e \bar{\nu}_\mu$ and dash-dotted light gray for the process $\mu^- \mu^+ \to Z XX$ with a subsequent decay $Z\to \nu_\mu \bar{\nu}_\mu$. The bottom panels show the ratio to the fixed-order calculation with $e\mu$ beams to reduce non-VBF topologies.}
	\label{fig:EWPDF_XX}
\end{figure}

Last, we analyze the Beyond the Standard Model (BSM) production of heavy scalars. As a simplified benchmark, we introduce a heavy scalar particle, denoted as $X$, with a mass set to 1 TeV and rescaled SM Higgs couplings (for examples of simplified models with electroweak di-boson resonances, cf.~\cite{Kilian:2014zja,Kilian:2015opv,Brass:2018hfw}). The invariant mass distribution of the $XX$ system is presented in Fig.~\ref{fig:EWPDF_XX}. While the EWA perfectly describes the process up to very high invariant masses, the EW PDF curves are systematically shifted downward. The associated scale uncertainty is small, as the process is dominated by longitudinal $W$ bosons. This example highlights potentially measurable resummation effects, and demonstrates an ideal scenario for the collinear approximation, which can successfully apply to probing BSM models, conveniently, e.g., when performing large scans over the parameter space.

\section{Summary and Conclusions}
\label{sec:concl}

At sufficiently high energies, collinear factorization becomes a valid approximation for generic multi-particle scattering and production processes, within the SM and beyond.  The approximation enables the resummation of large collinear logarithms, governed by DGLAP equations which incorporate all SM particles.  For instance, processes involving photon and gauge boson fusion are logarithmically enhanced compared to fermion-annihilation processes without radiation.   In this work, we have integrated the EW PDF description as originally introduced in Ref. \cite{Han:2020uid} into the automated calculation package and Monte Carlo generator \whizard\ and explored a number of representative processes that can be studied at a high-energy $\mu^+\mu^-$ collider, as well as other high-energy lepton colliders. 
The implementation includes the complete SM spectrum, and it accounts for specific properties of the electroweak interactions in the PDF context, namely helicity dependence, mass thresholds, and the scheme dependence of on-shell projections. The PDF approach is suitable for simulating effects of BSM physics.

We have studied both the low-energy regime where only QED and QCD partons are active, and the high-energy regime where the complete spectrum of SM particles appears. Below the electroweak scale, the EW PDF framework allowed us to compute inclusive mini-jet production in high-energy lepton collisions. We have shown that the production rates are large enough for mini-jet production to constitute a dominant SM background at future multi-TeV muon colliders (cf.\ Figs.~\ref{fig:EWPDF_colored_2D} and \ref{fig:EWPDF_colored_scale}). Its implementation in \whizard makes this calculation readily accessible, providing an out-of-the-box tool for (parton-level) mini-jet simulations in muon collider phenomenology studies.

For high-energy processes, we have focused on vector-boson-initiated hard scattering and analyzed the differences between a full-fledged EW PDF approach, fixed-order diagrammatic calculations with exact kinematics, and the EWA as a purely kinematic approximation.  The impact of the factorization scale on the results induces uncertainties, which we have taken into account in the numerical results. 

We have verified that the collinear framework offers a convenient description of heavy-particle production initiated by longitudinal gauge bosons, such as $HH$ and $t\bar{t}$ production. For partonic invariant masses above roughly 1\,TeV, both the EWA and the EW PDF approach agree well with the full matrix-element calculation, and they differ from each other by 5--20\% (Figs.~\ref{fig:EWPDF_HH} and~\ref{fig:EWPDF_ttbar_bands}). The approximation also performs well for several other observables, for example, the transverse momentum of the subleading final-state particle (Fig.~\ref{fig:EWPDF_HH_pT}). Interestingly, Higgs bosons at high transverse momentum are produced mainly through $W_T W_T$ fusion (Fig.~\ref{fig:EWPDF_HH_pol}). This contradicts the naive expectation that $W_0 W_0$ fusion dominates, as it does for the invariant-mass distribution. When transversely polarized bosons initiate the process, as happens partially for $t\bar{t}$ and predominantly for $\nu\bar{\nu}$ production, the scale uncertainties are larger, although they decrease in the high-$p_T$ region (Figs.~\ref{fig:EWPDF_ttbar_bands} and~\ref{fig:EWPDF_vv_bands}). For differential distributions, the $2H_T$ scale choice is preferable to $\sqrt{\hat{s}}/2$, and with it the agreement stays within $20\%$ even for neutrino-pair production. $WW$ production is more problematic because of large gauge cancellations, but the approximation remains reasonable once the phase space is restricted to realistic detector coverage (Fig.~\ref{fig:EWPDF_WW_cut}). Finally, as a simplified model of BSM physics, we have considered heavy-scalar pair production, where the collinear approximation works particularly well in the full range of invariant masses (Fig.~\ref{fig:EWPDF_XX}).

In summary, the implementation of the EW PDF formalism provides a first step towards the quantitative description of the ``EW jet regime'' where the complete SM gauge symmetry becomes manifest, and electroweak particles behave similarly to QCD jets.   The PDF approximation to exclusive hard processes in the \whizard\ Monte Carlo framework will eventually be supplemented by simulating additional electroweak radiation in a unified parton-shower approach, which we defer to future work. In addition, this framework also paves the way to full SM PDFs for the proton for FCC-hh.

\acknowledgments

We want to thank Stefano Frixione, Maximilian L\"{o}schner, Fabio Maltoni, Davide Pagani, Richard Ruiz, Panagiotis Stylianou, and Aleksander Filip Żarnecki for many fruitful discussions on the subject of the paper.
This work has been supported by the Deutsche Forschungsgemeinschaft (Germany) under Germany’s Excellence Strategy-EXC 2121 ``Quantum Universe''-390833306 and
-EXC 3107 533766364 ``Color meets Flavor'' and under Open Access Publication Funding-491245950, by the Deutsche Forschungsgemeinschaft under grant 396021762 - TRR 257, as well as by the National Science Centre (Poland) under the OPUS research project no. 2021/43/B/ST2/01778.
This article is based on work from the COST Actions COMETA (CA22130) and MLQC4FC (CA24146), supported by COST (European Cooperation in Science and Technology). 
WK would like to express his gratitude to the University of Pittsburgh (Pitt PACC) for support and hospitality, where part of this work was completed. 
YM acknowledges the support as a Postdoctoral Researcher of the Fonds de la Recherche Scientifique de Belgique (F.R.S.-FNRS), Belgium.
KM acknowledges the support of the Polish National Agency for Academic Exchange under the Bekker Programme (Grant No. BPN/BEK/2025/1/00156).
The work of KX at MSU is supported in part by the U.S. National Science Foundation under Grants No. PHY-2112829, PHY-2013791, and PHY-2310497.
The work of KX at TDLI/SJTU is also supported by the National Natural Science Foundation of China and by the State Key Laboratory of Dark Matter Physics.

\appendix

\section{Parton Numbering Scheme}
\label{app:partons}

Electroweak interactions are chiral; hence, the EW PDFs depend on chirality~\cite{Han:2020uid}.  Instead of just $2\times 4 +1=9$ entries for light quarks and gluons as in the QCD PDF table (or $11$ in a five-flavor scheme), the EW PDF implementation reserves 59 parton entries (as usual, color degrees are not counted separately)\footnote{Several PDF components are related to each other if the masses of the corresponding particles are neglected, e.g. $f_{u_L} = f_{c_L}$, $f_{u_R} = f_{c_R}$, etc.~\cite{Garosi:2023bvq}. To allow for maximal flexibility in defining the grids, these correlations are not imposed in the code.}:
\begin{itemize}
    \item 24 entries for quarks (6 flavors $\times$ 2 helicities for particles and antiparticles),
    \item 12 entries for charged leptons (3 flavors $\times$ 2~helicities for particles and antiparticles), 
    \item 6 entries for neutrinos (3 flavors for left-handed neutrinos and right-handed antineutrinos),
    \item 4 entries for massless gauge bosons (2 helicities for the photon and the gluon),
    \item 9 entries for massive gauge bosons (3 helicities for the $Z$, $W^+$ and $W^-$ bosons),
    \item 1 entry for the Higgs boson,
    \item 3 entries for the mixed gauge bosons ($\gamma/Z_{L/R}, H/Z_0)$.
\end{itemize}

The current LHAPDF format does not support parton helicity as a separate variable. For internal bookkeeping, we encode the PDG code together with helicity as a three-digit identifier that we use to identify and evaluate the pre-generated grids.  For example, 301 denotes the left-handed $s$ quark, while $-240$ stands for the longitudinal $W^{-}$ boson. However, in the externally visible event records we maintain the standard PDG numbering.

In Table~\ref{tab:PID}, we present the internal parton numbering scheme of \whizard, which can be used to interface with other PDF grids. Several remarks are in order. First, right-handed (left-handed) (anti)neutrinos are retained for reasons of symmetry, although the corresponding PDF entries in the grids are zero. Second, the current implementation retains the Higgs boson as a placeholder, but it is not included in the grids. Third, the current version does not use the three mixed parton states, $(\gamma/Z)_{L/R}$ and $H/Z_0$, to convolve with the corresponding (interference) matrix elements~\cite{Han:2020uid,Marzocca:2024ica}. The inclusion of these states, while contributing moderately to the final calculations, is in contradiction to the underlying default SM implementation in its broken phase and poses some conceptual challenges to the internal matrix-element generator, which has been left for further work within the \whizard collaboration. 

\begin{table}[]
    \centering
    \begin{tabular}{c||c|c|c|c|c|c||c|c|c|c|c|c}
    \hline\hline
        PID & -601 & -501 & -401 & -301 & -201 & -101 & 101 & 201 & 301 & 401 & 501 & 601    \\
          & $\bar{t}_R$ & $\bar{b}_R$ & $\bar{c}_R$ & $\bar{s}_R$ & $\bar{u}_R$ & $\bar{d}_R$ & $d_L$ & $u_L$ & $s_L$ & $c_L$ & $b_L$ & $t_L$    \\
         \hline
         PID & -602 & -502 & -402 & -302 & -202 & -102 & 102  & 202 & 302 & 402 & 502 & 602    \\
         & $\bar{t}_L$ & $\bar{b}_L$ & $\bar{c}_L$ & $\bar{s}_L$ & $\bar{u}_L$ & $\bar{d}_L$ & $d_R$ & $u_R$ & $s_R$ & $c_R$ & $b_R$ & $t_R$    \\ \hline\hline
        PID & -161 & -151 & -141 & -131 & -121 & -111 & 111 & 121 & 131 & 141 & 151 & 161    \\
         & $\bar{\nu}_{\tau,R}$ & $\tau^+_R$ & $\bar{\nu}_{\mu,R}$ & $\mu^+_R$ & $\bar{\nu}_{e,R}$ & $e^+_R$ & $e^-_L$ & $\nu_{e,L}$ & $\mu^-_L$ & $\nu_{\mu,L}$ & $\tau^-_L$ & $\nu_{\tau,L}$    \\
         \hline
        PID & -162 & -152 & -142 & -132 & -122 & -112 & 112  & 122 & 132 & 142 & 152 & 162    \\  
        & $\bar{\nu}_{\tau,L}$ & $\tau^+_L$ & $\bar{\nu}_{\mu,L}$ & $\mu^+_L$ & $\bar{\nu}_{e,L}$ & $e^+_L$ & $e^-_R$ & $\nu_{e,R}$ & $\mu^-_R$ & $\nu_{\mu,R}$ & $\tau^-_R$ & $\nu_{\tau,R}$    \\
         \hline\hline 
         PID & 211 & 221 & 231 & 241 & -241 & 261 & 230 & 240 & -240 & 250 & 270 \\
         & $g_{L}$ & $\gamma_{L}$ & $Z_{L}$ & $W^+_{L}$ & $W^-_{L}$ & $\gamma/Z_L$ & $Z_{0}$ & $W^+_{0}$ & $W^-_{0}$ & $H$ & $H/Z_0$  \\
         \hline
         PID & 212 & 222 & 232 & 242 & -242 & 262 \\
         & $g_{R}$ & $\gamma_{R}$ & $Z_{R}$ & $W^+_{R}$ & $W^-_{R}$ & $\gamma/Z_R$ \\
         \hline\hline
    \end{tabular}
    \caption{The PID scheme adopted in the EW PDF implementation of \whizard. The lower index corresponds to particle helicity.
    The PIDs for the right-handed neutrinos $\nu_{i,R}$ and the anti-particles $\bar{\nu}_{i,L}$ are explicitly present in the interface, but the contents are set to zero. Similarly, the mixed PDFs and the Higgs-boson one are unused in the current implementation.}
    \label{tab:PID}
\end{table}

\section{Example \texttt{SINDARIN} Input Files}
\label{app:sinEWPDF}
\texttt{SINDARIN} is \whizard's scripting language used for running the code with various options. Below, two example files to run \whizard for jet production (below the electroweak scale) and Higgs pair production (above the electroweak scale via $W$ boson fusion) in the EW PDF framework are presented:

\begin{itemize}
    \item \texttt{SINDARIN} file for $gg \to jj$:
\end{itemize}

\begin{lstlisting}
model = SM
ms = 0 
mc = 0
mb = 0
alias quark = u:d:s:c:b
alias QUARK = U:D:S:C:B
alias jet = quark:QUARK:g

sqrts = 10 TeV
beams = "mu+", "mu-" => lhapdf_ew
scale = eval 0.5*M [jet, jet] # Invariant mass of the di-jet over 2

process proc_jj = g, g => jet, jet

cuts = all M > 20 GeV [jet, jet]
   and all -2.44 < Eta < 2.44 [jet]
   and all Pt > (4 GeV + (sqrts / (3 TeV)) * 1 GeV) [jet] 

?alphas_is_fixed = false
?alphas_from_lhapdf = true

integrate (proc_jj)
\end{lstlisting}
\newpage
\begin{itemize}
    \item \texttt{SINDARIN} file for Higgs pair production:
\end{itemize}

\begin{lstlisting}
model = SM
sqrts = 10 TeV
beams = "mu+", "mu-" => lhapdf_ew
scale = eval 0.5*M [H, H] # Invariant mass of the Higgs pair over 2

process procHH = "W+", "W-" => H, H
integrate (procHH)
\end{lstlisting}

More information on available \texttt{cuts} and \texttt{selection} statements, as well as event generation, can be found in the \whizard manual.

\bibliographystyle{utphys}
\bibliography{ref.bib}
\end{document}